\documentstyle[12pt]{article}

\def\hybrid{\topmargin -20pt  \oddsidemargin 0pt
      \headheight 0pt   \headsep 0pt
     \textwidth 6.25in 
      \textheight 9.5in 
      \marginparwidth .875in
      \parskip 5pt plus 1pt   \jot = 1.5ex}

\hybrid

\begin{document}
\def\L{ {\cal L}}
\def\C{ {\cal C}}
\def\N{ {\cal N}}
\def\calE{{\cal E}}
\def\lin{{\rm lin}}
\def\Tr{{\rm Tr}}
\def\mxth{\mathsurround=0pt }
\def\xversim#1#2{\lower2.pt\vbox{\baselineskip0pt \lineskip-.5pt
x  \ialign{$\mxth#1\hfil##\hfil$\crcr#2\crcr\sim\crcr}}}
\def\simgr{\mathrel{\mathpalette\xversim >}}
\def\simle{\mathrel{\mathpalette\xversim <}}
\def\slash{\llap /}
\def\lagr{{\cal L}}

\newcommand{\ms}[1]{\mbox{\scriptsize #1}}
\renewcommand{\a}{\alpha}
\renewcommand{\b}{\beta}
\renewcommand{\c}{\gamma}
\renewcommand{\d}{\delta}
\newcommand{\th}{\theta}
\newcommand{\TH}{\Theta}
\newcommand{\pa}{\partial}
\newcommand{\g}{\gamma}
\newcommand{\G}{\Gamma}
\newcommand{\A}{\Alpha}
\newcommand{\B}{\Beta}
\newcommand{\D}{\Delta}
\newcommand{\e}{\epsilon}
\newcommand{\E}{\Epsilon}
\newcommand{\z}{\zeta}
\newcommand{\Z}{\Zeta}
\newcommand{\k}{\Kappa}
\newcommand{\K}{\Kappa}
\renewcommand{\l}{\lambda}
\renewcommand{\L}{\Lambda}
\newcommand{\m}{\mu}
\newcommand{\M}{\Mu}
\newcommand{\n}{\nu}
\newcommand{\X}{\Chi}
\newcommand{\R}{\Rho}
\newcommand{\s}{\sigma}
\renewcommand{\S}{\Sigma}
\renewcommand{\t}{\tau}
\newcommand{\T}{\Tau}
\newcommand{\y}{\upsilon}
\newcommand{\Y}{\upsilon}
\renewcommand{\o}{\omega}
\newcommand{\q}{\theta}
\newcommand{\h}{\eta}
\newcommand{\cmap}{{$\bf c$} map}
\newcommand{\Ka}{{K\"ahler}}
\renewcommand{\O}{{\Omega}}
\newcommand{\var}{\varepsilon}
\def\be{\begin{equation}}
\def\ee{\end{equation}}
\def\bea{\begin{eqnarray}}
\def\eea{\end{eqnarray}}

\newcommand{\ft}[2]{{\textstyle\frac{#1}{#2}}}
\newcommand  {\Rbar} {{\mbox{\rm$\mbox{I}\!\mbox{R}$}}}
\newcommand  {\Hbar} {{\mbox{\rm$\mbox{I}\!\mbox{H}$}}}
\newcommand {\Cbar}
    {\mathord{\setlength{\unitlength}{1em}
     \begin{picture}(0.6,0.7)(-0.1,0)
        \put(-0.1,0){\rm C}
        \thicklines
        \put(0.2,0.05){\line(0,1){0.55}}
     \end {picture}}}

\newcommand{\Al}{Alekseevski\v{\i}}
\newcommand{\eqn}[1]{(\ref{#1})}

\sloppy


\renewcommand{\thesection}{\arabic{section}}
\renewcommand{\theequation}{\thesection.\arabic{equation}}

\parindent0em

\begin{titlepage}
\begin{center}
\hfill KUL-TF-97/14\\
\hfill THU-97/09 \\
\hfill SWAT-97/150 \\[3mm]
\hfill {\tt hep-th/9707262}\\

\vskip 5mm

{\bf SPECIAL GEOMETRY IN HYPERMULTIPLETS}

\vskip .2in

{\bf J. De Jaegher$^a$, B. de Wit$^b$, B. Kleijn$^b$, 
S. Vandoren$^c$}\\ 
\vskip 1.2cm

$^a${\em Instituut voor theoretische fysica, Katholieke 
Universiteit Leuven,}\\
{\em B-3001 Leuven, Belgium}\\
{\tt Jeanne.DeJaegher@fys.kuleuven.ac.be}\\
$^b${\em Institute for Theoretical Physics, 
Utrecht University,}\\
{\em 3508 TA Utrecht, Netherlands}\\
{\tt  B.deWit@fys.ruu.nl, B.Kleijn@fys.ruu.nl} \\
$^c${\em Department of Physics, University of Wales Swansea,}\\
{\em  SA2 8PP Swansea, U.K.} \\
{\tt pysv@swansea.ac.uk}

\vskip .1in

\end{center}

\vskip .2in

\begin{center} {\bf ABSTRACT } \end{center}
\begin{quotation}\noindent 
We give a detailed analysis of pairs of vector and 
hypermultiplet theories with $N=2$ supersymmetry in four  
spacetime dimensions that are related by the (classical) mirror 
map. The symplectic reparametrizations of the special \Ka\ space 
associated with the vector multiplets induce corresponding 
transformations on the hypermultiplets. We construct the 
Sp(1)$\times$Sp($n$) one-forms  
in terms of which the hypermultiplet couplings are encoded and 
exhibit their behaviour under symplectic reparametrizations.
Both vector and 
hypermultiplet theories allow vectorial central charges in the 
supersymmetry algebra associated with integrals over the 
\Ka\  and hyper-\Ka\ forms, respectively. We show how these 
charges and the holomorphic BPS mass are related by the mirror 
map. 
\end{quotation}

\vskip 1cm

July 1997\\

\end{titlepage}
\vfill
\eject

\newpage

\section{Introduction}
In four spacetime 
dimensions with $N=2$ supersymmetry, there exist two inequivalent
matter supermultiplets. One is the vector multiplet, which 
comprises states of  
helicity $\pm 1$, $\pm\ft12$ and $0$, the other is the 
hypermultiplet with only states of helicity $\pm\ft12$ and 0. 
These supermultiplets appear in effective low-energy field 
theories for type-II string compactifications on a Calabi-Yau 
three-fold or for $N=2$ heterotic string compactifications on 
$K3\times T_2$. When considering superstring compactifications on 
a pair of mirror Calabi-Yau spaces \cite{mirror} the interesting 
phenomenon arises that the vector multiplets and the hypermultiplets in the 
four-dimensional effective action are interchanged. The same 
interchange is effected when compactifying type-IIA and  
type-IIB supergravity on the same Calabi-Yau manifold. This 
implies that, at least in string perturbation theory, the special 
\Ka\ and the quaternionic moduli spaces parametrized by the scalars of 
the vector multiplets and the hypermultiplets, are 
interchanged. In \cite{CecFerGir} this interchange was studied in 
detail, at the level of both supergravity and string 
theory, by reducing to three dimensions, where differences in 
helicity content of the multiplets no longer play a role. In that 
case the target space factorizes into two quaternionic spaces, 
corresponding to the two sets of inequivalent supermultiplets 
\cite{dWTNic}. The map from special \Ka\ to quaternionic 
manifolds was called the  
\cmap, because of its similarity to Calabi's construction for 
hyper-\Ka\ metrics on cotangent bundles of a \Ka\ 
manifold \cite{Calabi}. Through 
string duality \cite{stringdual}, there exist further relations between 
the vector multiplet and hypermultiplet sectors. 

In the rigid supersymmetry limit, a quaternionic manifold
reduces to a hyper-\Ka\ manifold. 
In this paper we study the hyper-\Ka\ manifolds that are in the 
image of the \cmap, but we  
expect that our results can be generalized rather 
straightforwardly to quaternionic manifolds in the context of an 
appropriate superconformal multiplet calculus. In this way we 
cover quite a large class of hyper-\Ka\ manifolds. While the form of 
the metric in `special coordinates' has been known for quite some 
time \cite{CecFerGir, 
HKLR}, we study the behaviour of various geometric quantities, 
such as the Sp(1)$\times$Sp($n$) one-forms in terms of which the 
hyper-\Ka\ manifold is defined \cite{BagWit}, under the 
diffeomorphisms induced by the symplectic  
reparametrizations in the underlying special \Ka\ space. Our hope 
is that, eventually, this will help us to constrain the 
perturbative string corrections for the hypermultiplets in 
type-II string compactifications, 
guided by what we know from the vector multiplet side. In other 
words, we intend to explore the consequences of special geometry for the 
corresponding systems of hypermultiplets.  

In practical applications hyper-\Ka\ and quaternionic manifolds are  
difficult to deal with because they are usually encoded in terms 
of a $(4n)$-dimensional metric, whose equivalence classes are 
provided by general diffeomorphisms. This in 
contradistinction to the special \Ka\ manifolds, which are 
conveniently encoded in terms of a holomorphic function 
\cite{function} and whose equivalence classes are described 
by a more restricted group of reparametrizations, due to 
supersymmetry and gauge invariance. These reparametrizations are associated with 
symplectic  matrices, which act on the (anti-)selfdual components of the 
field strenghts. Fixed points of these transformations 
correspond to invariances of the equations of motion. Such 
duality invariances have a long history in extended supergravity 
theories \cite{dual,dWVP}. As it turns out the description in 
terms of holomorphic functions and the symplectic 
reparametrizations are essential  
ingredients in the definition of `special geometry' 
\cite{special,FerStro}. They were important tools in the study of 
Calabi-Yau manifolds \cite{Cand}, nonperturbative phenomena 
in supersymmetric gauge theories \cite{SW}, as well as in studies of 
low-energy effective actions for vector multiplets arising from 
$N=2$ supersymmetric string compactifications and tests of string 
duality \cite{effective}. 

In the past,  important information on quaternionic  
manifolds was obtained via the \cmap. For instance, it was argued in  
\cite{CecFerGir,Cecotti} that the \cmap\ is closely related to the 
method employed in \cite{Aleks} for the classification of normal 
quaternionic spaces (i.e., quaternionic spaces that admit a 
solvable transitive group of isometries). In \cite{ssss} a 
general analysis was presented of the isometry structure of the 
quaternionic spaces in the  
image of the \cmap. With the exception of the quaternionic projective 
spaces, all normal quaternionic spaces are contained in the image of the 
\cmap. Part of the solvable algebra of isometries coincides with 
the duality invariances of the underlying special \Ka\ spaces. 
The symplectic reparametrizations of the special \Ka\ space now 
correspond to a subclass of the diffeomorphisms of the 
quaternionic space whose effect can be fully incorporated in 
the holomorphic function on which the quaternionic or hyper-\Ka\ 
metric depends.        

In view of these applications we will exhibit the effect of the 
symplectic reparametrizations on various quantities that play 
a role on the hyper-\Ka\ side. As alluded to above, among those 
are the  Sp(1)$\times$Sp($n$) one-forms, which we determine 
explicitly from the \Ka\ side. In order to do this we have to 
cast the results of \cite{BagWit} on the general hypermultiplet 
Lagrangians in a different form. Under the symplectic 
reparametrizations the hyper-\Ka\ forms transform covariantly. 
This means that the duality invariances on the \Ka\ side take the 
form of triholomorphic isometries of the hyper-\Ka\ manifold. 

We also consider the possible central charges that may be generated 
as surface terms in the anticommutator of the supersymmetry charges. 
It turns out that  
the vector multiplets generate the scalar and pseudoscalar 
charges associated with the holomorphic BPS mass and a vectorial 
central charge expressed in terms of the integral over the 
pull-back of the \Ka\ form. The hypermultiplets on the other hand 
only exhibit vectorial charges expressed as integrals over the 
pull-back of the hyper-\Ka\ forms. In three dimensions the central 
charges associated with these two multiplets can be related via 
the classical mirror map. 

We should add that our work has no direct implication for recent studies 
of nonperturbative supersymmetric gauge dynamics in 
three dimensions \cite{3dmirror}. In those studies one deals with effective 
actions based on a nonabelian gauge theory, where 
one of the three spatial dimensions is compactified to a circle 
of finite or zero length. However, in 
our work we start from a generic four-dimensional abelian gauge 
theory (which may be related to the effective action of some  
underlying supersymmetric gauge theory) without paying attention 
to its possible dynamic origin. Upon compactification of 
one dimension to a circle, this theory does not fully capture the 
dynamics of the underlying theory associated with the circle 
compactification. So within this setting we have to content 
ourselves with exploiting the  
relation between two classes of four-dimensional supersymmetric 
theories, based on vector multiplets and hypermultiplets, 
respectively. 

This paper is organized as follows. In section 2 we introduce the 
general action for vector multiplets and the corresponding 
symplectic reparametrizations. In section 3 we discuss the reduction of 
these theories to three dimensions and elucidate certain features 
relevant for finite compactification radius. Furthermore the 
geometry of the resulting hyper-K\"ahler target space is studied, 
including a classification of its isometries. In section~4 
we present a short derivation of the general supersymmetric action 
and transformation rules for hypermultiplets, in a 
slightly different setting than in \cite{BagWit}.   
In section~5 we discuss the emergence of an extra SU(2) symmetry 
group, contained in the automorphism group of the 
supersymmetry algebra, when descending to three dimensions. We 
give a detailed treatment of the classical mirror map and use it 
to determine explicit expressions for the one-forms in terms of 
which the hyper-\Ka\ manifold is defined, as well as other 
quantities of interest. These one-forms transform covariantly 
under the   
symplectic reparametrizations induced by the underlying \Ka\ 
geometry. We close with a discussion of the various central charges that 
may emerge in theories with vector multiplets and 
hypermultiplets and we  exhibit their relation under the mirror map.


\section{Vector multiplets}  
\setcounter{equation}{0}
We start by a discussion of the features that are  relevant for 
this paper of rigid $N=2$ supersymmetric systems in 
four dimensions consisting of vector multiplets. In particular we 
emphasize the symplectic reparametrizations of systems of abelian 
vector multiplets.
As is well known \cite{function}, the general supersymmetric 
Lagrangian for $N=2$ vector multiplets is encoded in  terms of a 
holomorphic function  
$F(X)$. Here the arguments $X^I$ ($I=1,2,\ldots,n$) denote the 
complex scalar fields of the $n$ vector multiplets. In the 
context of this paper we restrict ourselves to abelian vector 
multiplets. The corresponding Lagrangian can be read off from  
\cite{DWLVP} and, after elimination 
of auxiliary fields and Fierz reordering, is equal to
\bea
4\pi\, \lagr&\!=\!&
  i\Big(\pa_\mu F_I\,\pa^\mu \bar X^I-\pa_\mu \bar F_I\,\pa^\mu 
X^I\Big) \nonumber\\ 
  & & +\ft14 i\Big(F_{IJ}F^{-I}_{\mu\nu}F^{-J\,\mu\nu}
        -\bar F_{IJ}F^{+I}_{\mu\nu}F^{+J\,\mu\nu}\Big)\nonumber\\
  & & -\ft14 N_{IJ}\Big(\bar\O^{iI} \pa \slash\O_i^J
      +\bar \O^I_i\pa\slash\O^{iJ}\Big)
    -\ft14 i\Big(\bar \O_i^I\pa\slash F_{IJ} \O^{iJ} 
-\bar\O^{iI}\pa\slash \bar F_{IJ} \O_i^J\Big) \nonumber\\ 
  & & -\ft18 i \Big( F_{IJK}\bar\O_i^I\,\s^{\m\n} 
F_{\m\n}^{-J}\,\O_j^K\, \varepsilon^{ij}
      -\bar F_{IJK}\bar \O^{iI}\,\s^{\m\n} 
F_{\m\n}^{+J}\,\O^{jK}\,\varepsilon_{ij} \Big) 
      \nonumber\\
  & & +\ft1{96}i \Big(F_{IJKL} +i N^{MN}(2F_{MIK}F_{JLN} -\ft12 
F_{MIJ}F_{KLN}) \Big) \, 
\bar\O_i^I\s_{\m\n}\O_j^J \varepsilon^{ij} \,
\bar\O_k^K\s^{\m\n}\O_l^L \varepsilon^{kl} \nonumber \\
  & & -\ft1{96}i \Big(\bar F_{IJKL} -i N^{MN}(2\bar F_{MIK}\bar 
F_{JLN} -\ft12  \bar F_{MIJ}\bar F_{KLN}) \Big) \, 
\bar\O^{iI}\s_{\m\n}\O^{jJ} \varepsilon_{ij} \,
\bar\O^{kK}\s^{\m\n}\O^{lL} \varepsilon_{kl} \nonumber\\ 
  & & -\ft1{16} N^{MN} F_{MIJ} \bar F_{KLN}\,\bar\O^{iK}\O^{jL}\,
     \bar\O_i^I\O_j^J\,,
  \label{4dvlagr}
\eea
where we use the notation 
\be
N_{IJ} = -i F_{IJ} + i\bar F_{IJ} \,, \qquad N^{IJ}\equiv 
\big[N^{-1}\big]^{IJ}\,,
\ee
with $F_{I_1\cdots I_k}$ denoting the $k$-th derivative of $F$. 
The fermion fields $\O$ carry a chiral SU(2) index, $i,j,
\ldots=1,2$. The spinors with lower SU(2) index, $\O_i^I$, are of 
positive chirality, i.e. $\g^5 \O_i^I=  \O_i^I$; the spinors with 
upper SU(2) index, $\O^{iI}$, are of negative chirality. The  
tensors $F^{\pm}_{\m\n}$ are the (anti-)selfdual components of 
the gauge fields. In the free theory the holomorphic function $F(X)$ 
is quadratic, and its second derivatives determine the coupling 
constants $g_{IJ}$ and generalized theta angles $\theta_{IJ}$ 
according to  
\be
F_{IJ} = {\theta_{IJ}\over 2\pi} + i {4\pi\over g^2_{IJ}}\,.
\ee

The nonlinear sigma model contained in \eqn{4dvlagr} exhibits an 
interesting geometry. The complex scalars $X^I$ 
parametrize an $n$-dimensional target space with metric $g_{I\bar 
J}= N_{IJ}$. This is a \Ka\ space: its metric equals $g_{I\bar J} 
= \pa^2 K(X,\bar X)/\pa X^I\,\pa \bar X^J$,  with \Ka\ 
potential  
\be
K(X,\bar X) = i X^I\, \bar F_I(\bar X) -i\bar X^I \, F_I(X) 
\,.  \label{Kpotential}
\ee
The resulting geometry is known as {\it special geometry}. 
Nonvanishing components of the Levi-Civit\'a connection and the 
curvature tensor are given by   
\bea
\G^{I}_{JK} &\equiv& g^{I\bar L} \pa_J g_{K\bar L} = -i N^{IL} F_{JKL} 
\,,\nonumber \\ 
R^I{}_{JK}{}^L &\equiv& g^{L\bar L} \,\pa_{\bar L}\G^I_{JK}= 
- N^{IP} N^{LQ} N^{MN}\, \bar F_{PQM}  \, F_{NJK}\,.  
\label{specialconcurv}
\eea
It is possible to choose different coordinates and view the 
$X^I$ as holomorphic `sections' $X^I(z)$ \cite{CdAF}. As it is 
straightforward to cast our results in such a 
coordinate-independent form, we keep writing  them in terms of 
the $X^I$, which are sometimes called {\it special} 
coordinates\footnote{%
   Special coordinates are singled out by supersymmetry. They 
   are the lowest component of an N=2  chiral reduced multiplet.   
   Strictly speaking the term `special geometry' was proposed for 
   systems of vector multiplets with {\it local} supersymmetry 
   \cite{special}. To make a distinction one occasionally uses 
   the term  `rigid' special geometry.}. %

We also record the supersymmetry transformation rules for the 
vector multiplet components (after elimination of the auxiliary 
fields),
\bea
  \delta X^I &=& \bar \e^i\O_i^I,\nonumber\\
  \delta A_\mu^I &=& \varepsilon^{ij}\,\bar \e_i\g_\mu\O_j^I
    +\varepsilon_{ij}\,\bar \e^i\g_\mu\O^{jI},  \label{4dvsusy}\\
\delta\O_i^I +\G^I_{JK} \, \d X^J\,\O_i^K &=& 2\pa \slash 
X^I\e_i-i \varepsilon_{ij}\s^{\m\n}\e^j \,  N^{IJ} 
{\cal G}^-_{\m\n J}    + \ft12 i N^{IJ} \bar 
F_{JKL} \bar\O^{kK} \O^{lL} \, \varepsilon_{ik}\varepsilon_{jl} \,
\e^j \,, \nonumber
\eea
where $\G$ denotes the \Ka\ connection and ${\cal G}_{\m\n I}^-$ is an 
anti-selfdual tensor defined by
\be
{\cal G}^-_{\m\n I} = i N_{IJ}F^{-J}_{\m\n}  -\ft14 F_{IJK} \bar\O 
^J_i\s_{\m\n}\O^K_j\,\varepsilon^{ij} \,. \label{defcG}
\ee
The significance of the tensor \eqn{defcG} and of the particular 
form of the spinor transformation in \eqn{4dvsusy}, will be discussed 
shortly. The supersymmetry transformation parameters, $\e^i$ and 
$\e_i$, are of positive and negative chirality, respectively. They 
transform as doublets under the chiral SU(2)$_{\rm R}$ group, 
which belongs to the automorphism group of the supersymmetry algebra. 
Observe that both the Lagrangian \eqn{4dvlagr} and transformation rules 
\eqn{4dvsusy} are consistent with respect to SU(2)$_{\rm R}$, but 
not, in general, with respect to the U(1)$_{\rm R}$ subgroup of 
the automorphism group.

It is possible for two different functions $F(X)$ to describe the 
same theory. The equivalence is provided by symplectic 
reparametrizations associated with the group Sp$(2n;{\bf Z})$. 
The discrete nature of this group is tied to nonperturbative 
effects, as the lattice of electric and magnetic charges should 
be left invariant. At the perturbative level, the group is 
Sp$(2n;{\bf R})$. The equivalence follows from rotating the 
Bianchi identities, $\partial^\mu \big(F^{+} 
-F^{-}\big){}^I_{\m\n} =0$, and the field equations for the 
vector fields, $\partial^\mu \big(G^+ -G^-\big){}_{\m\n I} =0$, 
by means of a real symplectic $(2n)$-by-$(2n)$ matrix. 
The tensors $G^\pm_{\m\n I}$ are obtained from the Lagrangian 
\eqn{4dvlagr}, and read
\be 
G^-_{\m\n I} =  F_{IJ}\,F^{-J}_{\m\n}  -\ft14 F_{IJK} \bar\O 
^J_i\s_{\m\n}\O^K_j\,\varepsilon^{ij} \,, \label{defG}
\ee
while $G^+_{\m\n I}$ is related to $G^-_{\m\n I}$ by complex 
conjugation. The symplectic rotation between equations of motion 
and Bianchi identities is induced by 
\be
\pmatrix{F^{\pm I}_{\mu\nu}\cr  G^\pm_{\mu\nu I}\cr} 
\longrightarrow  \pmatrix{U&Z\cr W&V\cr} \pmatrix{F^{\pm 
I}_{\mu\nu}\cr  G^\pm_{\mu\nu I}\cr}\,,\label{FGdual}
\ee
where $U^I_{\,J}$, $V_I^{\,J}$, $W_{IJ}$ and $Z^{IJ}$ are 
constant real  $n\times n$ submatrices, subject to certain 
constraints such that the total matrix is an element of 
Sp$(2n;{\bf R})$. Consistency with \eqn{defG} requires the 
symmetric tensor $F_{IJ}$ to change as ${F}_{IJ} \to (V_I{}^K 
{F}_{KL}+ W_{IL} )\, \big[(U+ Z{F})^{-1}\big]^L{}_J$. 
This is achieved by the transformations of the scalar fields, 
implied by 
\be
\pmatrix{X^{I}\cr  F_{I}\cr} \longrightarrow  \pmatrix{\tilde 
X^I\cr\tilde F_I\cr}=
\pmatrix{U&Z\cr W&V\cr} \pmatrix{X^{I}\cr  F_I\cr}\,. 
\label{transX}
\ee
In this transformation we include a transformation of $F_I$. 
Because the transformation is symplectic, one can show that the new 
quantities $\tilde F_I$ can be written as the derivatives of a 
new function $\tilde F(\tilde X)$. The new equations of motion 
after performing \eqn{FGdual} and \eqn{transX} then follow 
straightforwardly from the Lagrangian based on $\tilde F$ 
(provided we perform suitable redefinitions of the spinor fields, 
which we will specify shortly).  

It is possible to integrate \eqn{transX} and determine the new 
function $\tilde F$,  
\bea
\tilde F(\tilde X) &=& F(X)-{\textstyle{1\over2}}X^I F_I(X) 
\label{newfunction} \\
&& + {\textstyle{1\over2}} \big(U^{\rm T}W\big)_{IJ}X^I X^J 
+{\textstyle{1\over2}}
\big(U^{\rm T}V + W^{\rm T}  Z\big)_I{}^J X^IF_J 
+{\textstyle{1\over2}} \big(Z^{\rm T}V\big){}^{IJ}F_I F_J \,,   
 \nonumber
\eea
up to a constant and terms linear in the $\tilde X^I$ (which give 
no contribution to the Lagrangian \eqn{4dvlagr}). In the coupling to 
supergravity, where the function must be homogeneous of second 
degree, such terms are excluded.\footnote{%
   The terms linear in $\tilde X$ in \eqn{newfunction} are 
   associated with constant translations in $\tilde 
   F_I$ in addition to the symplectic rotation shown in 
   \eqn{transX}. Likewise one may introduce constant shifts in
   $\tilde X^I$. Henceforth we ignore these shifts. Constant 
   contributions to $F(X)$ are always irrelevant. We note also 
   that terms quadratic in the $X^I$ with real coefficients 
   correspond to total divergences in the action.} %
Obviously $F(X)$ does not transform as a function. Such 
quantities turn out to be rare. Examples are the holomorphic 
function $F(X)-\ft12 X^IF_I(X)$ and the \Ka\ potential 
\eqn{Kpotential}. For a discussion of this, we 
refer to \cite{DW,hograv}. In practical situations the 
expression \eqn{newfunction} is not always useful, as it 
requires substituting $\tilde X^I$ in terms of $X^I$, or vice 
versa. When $F$ remains the same, $\tilde F(\tilde X) = F(\tilde 
X)$,  the theory is {\it invariant} under the 
corresponding transformations. These invariances are often called 
duality invariances and they have been studied extensively in the 
context of extended supergravity theories \cite{dual,dWVP}. 
The space of inequivalent couplings of $n$ 
abelian vector supermultiplets is equal to the space of 
holomorphic functions of $n$ variables, divided by the ${\rm 
Sp}(2n;{\bf R})$ group. This group does not act freely on the space 
of these functions. There are  fixed points whenever the 
equations of motion exhibit duality symmetries. It is 
not easy to find solutions of $\tilde F(\tilde X) = F(\tilde 
X)$, unless one considers infinitesimal transformations. In that 
case the condition reads \cite{dWVP}
\be
C_{IJ}\,X^IX^J - 2 B^I{}_J\,X^JF_I + D^{IJ}\,F_IF_J= 0\,,
\ee
where the constant matrices $B^I{}_J$, $C_{IJ}$ and $D^{IJ}$ 
parametrize the infinitesimal form of the ${\rm Sp}(2n;{\bf 
R})$ matrix, according to $U\approx{\bf 1} +B$, $V\approx {\bf 1} 
-B^{\rm T}$, $W\approx C$ and $Z\approx -D$.
For finite transformations, a more convenient method is to verify  
that the substitution $X^ I\to \tilde X^I$ into the derivatives 
$F_I(X)$ correctly induces the symplectic transformations on the 
`periods' $(X^I,F_J)$. 

It is convenient to employ quantities that transform as tensors 
under symplectic reparametrization. Before considering some such 
tensors let us introduce the following notation,
\bea
{\partial\tilde X^I\over\partial X^J}\,&\equiv& \,{\cal 
S}^I{}_{\!J}(X)
= U^I{}_{\!J} +Z^{IK}\,F_{KJ}(X) \,, \nonumber\\
{\cal Z}^{IJ}(X) \,&\equiv& \,[{\cal S}^{-1}(X)]^I{}_K\, Z^{KJ} 
\,. \label{symplv}
\eea
The holomorphic quantity ${\cal Z}^{IJ}$ is symmetric in $I$ and $J$, because 
$Z\,U^{\rm T}$ is a symmetric matrix as a consequence of the fact 
that $U$ and $Z$ are submatrices of the symplectic matrix 
indicated in \eqn{FGdual}. 

After these definitions we note the following transformation 
rules,  
\bea
\tilde{F}_{IJ} &=& (V_I{}^K {F}_{KL}+ W_{IL} )\,
\big[{\cal S}^{-1}]^L{}_J\,, \nonumber\\ 
\tilde N_{IJ} &=& N_{KL}\, \big[\bar{\cal 
S}^{-1}\big]^K{}_{\!I}\,  \big[{\cal S}^{-1}\big]^L{}_{\!J}\,, 
\nonumber\\ 
\tilde N^{IJ} &=& N^{KL}\, \bar{\cal 
S}^I{}_{\!K}\,{\cal S}^J{}_{\!L}\,, \nonumber \\
\tilde F_{IJK} &=& F_{MNP}\, \big[{\cal S}^{-1}\big]^M{}_{\!I} \,
    \big[{\cal S}^{-1}\big]^N{}_{\!J}\, \big[{\cal 
S}^{-1}\big]^P{}_{\!K} \,, \nonumber\\
\tilde \O^I_i &=& {\cal S}^I{}_{J}\, \O_i^J\,, \qquad \tilde 
\O^{iI} \,= \,\bar{\cal S}^I{}_{J}\, \O^{iJ}\,. \label{sympltr}
\eea
The first three quantities do not remain manifestly symmetric in $I,
J$, but this symmetry is preserved   
owing to the symplectic nature of the transformation. The \Ka\ 
connection transforms as a mixed tensor but also acts as a 
connection for symplectic reparametrizations, as follows from 
\bea
\tilde \G^I_{JK} &=& \bar {\cal S}^I{}_L\, \G^L_{MN} \, [{\cal 
S}^{-1}]^M{}_J\, [{\cal S}^{-1}]^N{}_K \nonumber \\
&=& -\pa_M{\cal S}^I{}_N \,[{\cal S}^{-1}]^M{}_J\, [{\cal 
S}^{-1}]^N{}_K +{\cal S}^I{}_L\, \G^L_{MN}\, [{\cal 
S}^{-1}]^M{}_J\, [{\cal S}^{-1}]^N{}_K \,.
\eea

{}From the field strengths $F^{\pm I}$ and $G^\pm_I$ we can 
construct tensors that transform as symplectic vectors. An 
example is the tensor ${\cal G}$ that we defined in \eqn{defcG}, 
which follows from\footnote{%
   Replacing $\bar F_{IJ}$ by $F_{IJ}$ in \eqn{defcG2} leads to a 
   purely fermionic expression, which is also symplectically 
   covariant.}%
\be
{\cal G}^-_{\m\n I} = G^-_{\m\n I} - \bar F_{IJ}F^{-J}_{\m\n} \,,
\label{defcG2}
\ee
upon substitution of \eqn{defG}. This particular combination 
transforms under symplectic reparametrizations as
\be
\tilde {\cal G}^-_{\m\n I}= {\cal G}^-_{\m\n I}\,[\bar{\cal 
S}^{-1}]^J{}_I \,.
\ee
With this result one can verify that the spinor transformation 
rule in \eqn{4dvsusy} is manifestly covariant under symplectic 
reparametrizations. The same is true for the supersymmetry 
variation of the scalar field, but not for the variation of the 
vector field. This is not surprising, because the symplectic 
reparametrizations are not defined for the gauge fields. The 
reader may also verify that the Lagrangian \eqn{4dvlagr} is  
invariant under symplectic reparametrizations, but only up to 
terms proportional to the equations of motion  
of the vector fields.


\section{Reduction to three spacetime dimensions}
\setcounter{equation}{0}

In this section we reduce the general Lagrangian \eqn{4dvlagr} 
for (abelian) vector multiplets to three spacetime dimensions. 
This is done by compactifying one of the spatial dimensions (say, 
the one parametrized by $x^3$) on a circle with 
radius $R$ and suppressing all the modes that depend nontrivially 
on $x^3$. The four-dimensional gauge fields then decompose into 
three-dimensional gauge fields $A^I_\m$ and additional scalar 
fields $A^I\equiv A^I_3$. If we impose the Bianchi identity in three
dimensions through addition of a Lagrange multiplier term proportional
to $B_I\e^{\m\n\rho}\partial_\m F^I_{\n\rho}$ and
integrate out the field strength, the degrees of freedom of the 
four-dimensional gauge field are captured in the two scalars $A^I$ 
and $B_I$.  

Before turning to more explicit results we deal with the 
consequences of the dimensional reduction for the fermions, 
which, in four spacetime dimensions, are  
four-component Majorana spinors. When reducing to three spacetime 
dimensions, every spinor decomposes into two two-component 
spinors. In order to discuss this systematically one decomposes 
the Clifford algebra of the gamma matrices in four dimensions 
into two mutually {\it commuting} Clifford algebras: one is the
algebra generated by the gamma matrices appropriate to three 
dimensions and the second one is the algebra generated by $\g^3$. 
This is accomplished by defining 
\be
\g^\m = \g_{(4)}^\m\tilde \g\,, \quad(\m=0,1,2)
\ee
where 
\be
\tilde \g\equiv -i \g^3\g^5\,,
\ee
so that $\g^1\g^2\g^0$ is proportional to the identity matrix. 
This implies that the Clifford algebra generated by these 
three-dimensional gamma matrices acts on the two-component 
spinors in equivalent representations. The gamma 
matrices $\g^3$ and $\g^5$ coincide with their  
higher-dimensional expressions: $\g^3= \g^3_{(4)}$ and $\g^5= 
\g^5_{(4)}$. The matrices $\tilde \g$, $\g^3$ and $\g^5$ 
{\it commute} with the three $\g^\m$. An observation that will be 
relevant later on, is that  
$\hat\s^1\equiv \g^3$, $\hat\s^2\equiv\g^5$ and 
$\hat\s^3\equiv\tilde \g$ form  an $su(2)$ algebra: 
$\hat\s^1\,\hat\s^2=i\hat\s^3$. 

Because the Dirac conjugate of a spinor involves the matrix 
$\g^0$, it will acquire an extra factor $\tilde \g$ as compared 
to the four-dimensional definition. Correspondingly we absorb a 
factor $\tilde\g$ into the three-dimensional  
charge-conjugation matrix $C$. With this definition we  
have the following identities
\be
C\,\g^\m C^{-1}=- \g^{\m\rm T},\quad C\,\g^3 C^{-1}= \g^{3\rm T},
\quad C\,\tilde\g C^{-1}= \tilde\g^{\rm T},\quad C\,\g^5 C^{-1}= 
-\g^{5\rm T}\,.
\ee
It is possible to choose $C$ such that it commutes with $\g^3$, 
$\tilde\g$ and $\g^5$.

The material of this section is organized in three subsections. 
First we discuss the actual reduction leading to a supersymmetric 
nonlinear sigma model in three dimensions. Then we elucidate the 
geometrical aspects of the target space. Finally, in a last 
subsection, we discuss the isometry structure of the target 
space. 

\subsection{The reduction}
Now we turn to the Lagrangian of the compactified theory. After converting
the three-dimensional gauge field into a scalar field, the terms in 
the Lagrangian \eqn{4dvlagr} that contain the field strengths, 
are replaced by 
\bea 
&& -\ft14{i}\Big(\bar F_{IJ}\,W_\mu^IW^{J\mu}
        -F_{IJ}\,\bar W_\mu^I\bar W^{\mu J}\Big)\nonumber\\
  & & -\ft18{i} \Big( F_{IJK}\bar \O_i^I\,\bar W^J_\mu \g^\mu \g_3
   \O_j^K\varepsilon^{ij} 
      - \bar F_{IJK}\bar \O^{iI}\, W^J_\mu\g^\mu 
\g_3\O^{jK}\varepsilon_{ij}\Big)\,, 
  \label{3dvlagr'}
\eea
where
\bea
  W_\mu^I &=& 2iN^{IJ}(\pa_\mu B_J-F_{JK}\,\pa_\mu A^K)\nonumber\\
   & &+\ft14{i}N^{IJ}\Big(\bar F_{JKL}\,\bar 
\O^{iK}\g_\mu\g_3\O^{jL}\varepsilon_{ij}
      +F_{JKL}\,\bar 
\O_i^K\g_\mu\g_3\O_j^L\varepsilon^{ij}\Big)\,.
  \label{Wmu}
\eea
Substituting this into \eqn{3dvlagr'} and combining with the other 
terms of the Lagrangian \eqn{4dvlagr} yields
\bea
4\pi\, \lagr & = & i \Big(\pa_\m F_I \,\pa^\m \bar X^I - \pa_\m \bar F_I \,
\pa^\m  X^I\Big)   - 
N^{IJ} (\partial_{\mu}B_I-F_{IK}\partial_{\mu}A^K)
(\partial^{\mu}B_J-  \bar{F}_{JM}\partial^{\mu}A^M) \nonumber\\ 
  & &{} -\ft14 N_{IJ}\Big(\bar\O^{iI} \pa \slash\O_i^J
      +\bar \O^I_i\pa\slash\O^{iJ}\Big) \nonumber\\ 
&& {}-\ft14 i F_{IJK} \Big(\bar \O_i^I\pa\slash X^J \O^{iK}  
- i \bar{\Omega}^I_i N^{JL}(\pa\slash B_L-\bar{F}_{LM}
\pa\slash A^M)\gamma_3 \Omega^K_j\varepsilon^{ij}\Big)  \nonumber\\   
& & {}+\ft{1}{4}i\bar F_{IJK} \Big(\bar\O^{iI}\pa\slash \bar X^J 
\O_i^K 
+i\bar{\Omega}^{iI} N^{JL} (\pa\slash B_L-F_{LM}\pa\slash 
A^M)\gamma_3 \Omega^{jK}\varepsilon_{ij}\Big)     \label{3dvlagr}  \\
& &{}+\ft{1}{96}i
\Big(F_{IJKL}+ 3 iN^{MN}F_{M(IJ}F_{KL)N}\Big) \bar{\Omega}^I_i\,
\gamma_3 \gamma_{\mu}\Omega^J_j \varepsilon^{ij}\bar{\Omega}^K_k
\gamma_3 \gamma^{\mu}\Omega^L_l \varepsilon^{kl} \nonumber\\
& &{}-\ft{1}{96}i
\Big(\bar{F}_{IJKL}-3iN^{MN}\bar{F}_{M(IK}\bar{F}_{JL)N}\Big) 
\bar{\Omega}^{Ii}\gamma_3 \gamma_{\mu}\Omega^{Jj} 
\varepsilon_{ij}\, 
\bar{\Omega}^{Kk}\gamma_3 \gamma^{\mu}\Omega^{Ll} 
\varepsilon_{kl} \nonumber \\
&& {}-\ft1{48} N^{MN} F_{MIJ}\bar F_{KLN}\, 
\Big(2\bar\O^I_i\g_\m\O^{iK} \, \bar\O^J_j\g^\m\O^{jL}  +
\bar\O_i^I\g_\m\g_3 \O^J_j \varepsilon^{ij} \,\bar 
\O^{kK}\g_\m\g_3  \O^{lL} \varepsilon_{kl} \Big)
\,,  \nonumber 
\eea
where we have suppressed a factor $2\pi R$ corresponding to the 
integration over the compactified coordinate $x^3$. Observe that 
the Lagrangian remains manifestly invariant under SU$(2)_{\rm 
R}$. Note also that we keep the fermion fields in their original
four-dimensional form, i.e. they are doublets of $\ft12(1\pm 
\g^5)$ projections of four-dimensional  Majorana spinors. Only 
the definition of the Dirac conjugate has been changed in accord 
with the rules obtained above. 

The above Lagrangian is invariant under the following 
supersymmetry transformations, 
\bea
  \delta X^I &=& -i \bar\e^i \g_3 \O_i^I  \,,       \nonumber\\
  \delta A^I &=& i\varepsilon^{ij}\bar \e_i \O_j^I
    - i\varepsilon_{ij}\bar\e^i\O^{jI}  \,,           \nonumber\\
  \delta B_I &=&  iF_{IJ}\varepsilon^{ij}\bar \e_i\O_j^J
   -i \bar F_{IJ} \varepsilon_{ij}\bar \e^i\O^{jJ} \,,  \nonumber\\
\d\O^I_i &=& 2i \pa\slash X^I \g_3 \e_i + 2 N^{IJ}(\pa\slash B_J- 
\bar F_{JK} \pa\slash A^K)\varepsilon_{ij} \e^j \nonumber \\
&&  
+i N^{IJ}\d F_{JK}\,\O^K_i -N^{IJ}\bar F_{JKL} N^{KM}(\d B_M- 
F_{MN}\d A^N) \varepsilon_{ij}\g_3 \O^{Lj}\,,\nonumber \\
\d\O^{Ii}&=& -2i \pa\slash \bar X^I \g_3 \e^i + 2 N^{IJ}(\pa\slash B_J- 
F_{JK} \pa\slash A^K)\varepsilon^{ij} \e_j \nonumber\\
&&   -i N^{IJ}\d \bar F_{JK}\,\O^{Ki} - N^{IJ} F_{JKL} N^{KM}(\d B_M- 
\bar F_{MN}\d A^N) \varepsilon^{ij}\g_3 \O_j^L \,.
\label{3dvsusy}
\eea
Under symplectic reparametrizations $(A,B)$ transform as a 
symplectic pair, just as the field strengths (cf. \ref{FGdual}). 
{}From $(A^I,B_I)$
we can construct a complex scalar 
\be
Y_I = B_I - F_{IJ}A^J\,, \label{defY}
\ee 
which transforms as a (co)vector under symplectic reparametrizations, 
\be
  \tilde Y_I= Y_J\,[{\cal S}^{-1}]^J{}_I \,. \label{Ytrans}
\ee
The supersymmetry transformation rule for $Y_I$ equals 
\be
  \delta Y_I- \G_{JI}^K \,\delta X^J \,Y_K = 
-N_{IJ}\varepsilon_{ij}\bar \e^i\O^{jJ} +i F_{IJK}\,  N^{JL}\,
\bar Y_L \, \bar\e^i\g_3\O^K_i   \,,
\ee
where the left-hand side takes the form of a symplectically 
covariant variation, while the right-hand side is explicitly 
symplectically covariant. Observe that $X^I$ and $Y_I$ 
all transform holomorphically, i.e., their supersymmetry 
variations are proportional 
to $\bar\e^i$ and not to $\bar\e_i$. This observation will be 
relevant in subsection~3.3. All supersymmetry 
variations take a symplectically covariant form, as follows from 
using the transformation properties given in section~2.

After the dualization of the vector to scalar fields, 
the symplectic reparametrizations can be applied to the  
equations of motion or directly to the Lagrangian. These 
reparametrizations express the fact that  
the theory retains its form under certain diffeomorphisms, 
provided that we simultaneously change the function $F(X)$. As 
with general diffeomorphisms, this is not an invariance 
statement, but it characterizes the equivalence classes of the 
theory as encoded in functions $F(X)$. Henceforth we 
will use the term `symplectically invariant' to indicate that 
quantities retain their form under the combined effect of a 
certain diffeomorphism and a change of the function $F(X)$. The 
Lagrangian \eqn{3dvlagr} is symplectically invariant. In 
particular we note that the four-fermion terms are proportional to 
either the special \Ka\ curvature or to the symmetric tensor
\be
C_{IJKL} = F_{IJKL} +3iN^{MN}\,F_{M(IJ}\,F_{KL)N}\,. 
\label{Ctensor}
\ee
Both tensors are symplectically covariant. The latter tensor 
vanishes for a symmetric \Ka\ space (defined by the condition 
that the curvature tensor is covariantly constant). As mentioned 
above, the above supersymmetry transformations are 
covariant under symplectic reparametrizations. 
 
If the function $F(X)$ describes an effective four-dimensional 
gauge theory, based on 
charged fields which have been integrated out, then 
the $\theta$-angles are defined up to shifts by  
$2\pi$ (at the nonperturbative level). Consequently the quantity 
$F_{IJ}$ is only defined up to an additive integer-valued 
matrix.\footnote{%
  In principle the integers are multiplied by a certain 
  constant determined by the embedding of the corresponding 
  U(1) group into the nonabelian gauge group of the underlying 
  field theory. This constant is set to unity. } %
{}From this observation it follows that, after compactifying on a 
circle, we must identify $B_I$ with $B_I$ plus an integer times 
$A^I$. Furthermore the fields 
$A^I$ are only defined up to an integer times $1/R$, as a 
consequence of four-dimensional gauge transformations with 
non-trivial winding around the compactified direction.\footnote{%
  For simplicity, we have set the $I$-th elementary charge 
  equal to unity.} %
Therefore, consistency requires that also $B_I$ is 
defined up to an integer times $1/R$. At the perturbative 
level the corresponding invariance is realized by continuous 
transformations as can be seen from \eqn{3dvlagr}, which is 
invariant under $F_{IJ}\to F_{IJ} + c_{IJ}$ and $B_I\to 
B_I+ c_{IJ} A^J$, where the constants $c_{IJ}$ constitute an 
arbitrary real symmetric tensor. These transformations correspond 
to the  
continuous Peccei-Quinn symmetries and are consistent with the 
transformations induced by the symplectic reparametrizations 
of the underlying vector-multiplet theory. Note that these 
transformations do not presuppose invariance under continuous 
shifts of the fields $A_I$, which, at finite $R$, do not 
represent a symmetry at the perturbative level. It is here that 
our approach fails to capture the dynamical effects 
associated with the compactification, just because we take $F(X)$ 
from a four-dimensional setting. This has no direct bearing on 
the fact that the target 
space parametrized by the $(A^I,B_I)$ fields  
constitutes a torus $T^{2n}$, whose periodicity lattice is 
in fact directly related to the lattice of dyonic 
charges.  Globally the full space is a fibre bundle over a 
special \Ka\ manifold with fibre $T^{2n}$. In the limit 
$R\rightarrow 0$, the torus decompactifies to ${\bf R}^{2n}$. 

Let us discuss some properties of the torus at a given point $X$ 
in the special-\Ka\ moduli space. First we determine the volume of 
$T^{2n}$, which turns out to be independent of $X$. To see 
this one integrates the square root of the determinant of the $(A^I, 
B_I)$ metric given in \eqn{3dvlagr} over the 
torus. Including the factor $4\pi$ from the left-hand side of 
\eqn{3dvlagr} and the factor $2\pi R$ from the integration over 
the compactified coordinate $x^3$, we find
\be
V(T^{2n}) = (4R)^{-n} \,.
\ee
Secondly, consider the invariant
lengths of cycles $\gamma(X): t\mapsto (X^I;A^I(t),B_I(t))$, 
which depend on the point $X$ in the special-\Ka\ moduli space. 
For the cycles $\gamma_{A^I}$ and
$\gamma_{B^I}$ in the $A^I$ and $B^I$ directions, these lengths 
are equal to 
\be
  \ell_{A^I}(X) = \frac{1}{R}\sqrt{(FN^{-1}\bar{F})_{II}}\, ,\qquad
  \ell_{B_I}(X) = \frac{1}{R}\sqrt{(N^{-1})^{II}}\,.
\ee
When $X^I$ approaches a point where the \Ka\ metric becomes 
singular, one of the cycles $(\gamma_{A^I},\gamma_{B_I})$ shrinks 
to zero while the other one grows to infinite length.  

At this point it is tempting to identify the torus at $X$ with the 
Jacobian variety of an auxiliary Riemann surface ${\cal M}_X$ that 
underlies the four-dimensional nonperturbative dynamics of a gauge theory 
in the Coulomb phase \cite{SW}. Its effective action takes the 
form of \eqn{4dvlagr} and the abelian vector multiplets are then 
associated with the Cartan  
subalgebra of the underlying gauge group. Singularities in the 
effective action associated with the emergence of massless states 
correspond to a pinching of the auxiliary Riemann surface  ${\cal 
M}_X$ which in turn leads to a degeneration of its Jacobian 
variety.\footnote{%
  In a full three-dimensional treatment, it is possible that 
  nonperturbative effects associated with  
  monopoles wrapping around the circle smooth out some of these 
  singularities; see the first reference of \cite{3dmirror}.} %
According to these arguments one may conclude that the complex
scalars $Y_I$ take their values in the (rescaled) Jacobian, 
\be
  J({\cal M}_X) = {\bf C}^n / L_X,\qquad L_X = 
\bigg\{\frac{1}{R} 
    \Big(m_I-F_{IJ}(X)\,n^J\Big)\bigg |\, m_I, n^I \in {\bf Z} \bigg\},
\ee
where we identify the second derivative of $F(X)$ with the
intersection matrix $\tau$ of the Riemann surface ${\cal M}_X$.  
The target space parametrized by all the scalar fields thus coincides 
with the holomorphic Sp$(2n;{\bf Z})$ bundle of Jacobian varieties over 
the moduli space of auxiliary Riemann surfaces, with metric as 
given in \eqn{3dvlagr} and transitions functions prescribed by 
the monodromies of the moduli space.

%
\subsection{Geometric features and symplectic transformations}
A number of geometric features of the target space associated 
with the metric defined in the Lagrangian \eqn{3dvlagr}  
deserves further attention. Note that, as a 
three-dimensional model, we are dealing with four independent  
supersymmetries. Therefore the target space must be a hyper-\Ka\ 
manifold \cite{AGF}, which, in the case at hand, is completely 
determined by the 
holomorphic function $F(X)$. Some of the properties of the 
special \Ka\ space are inherited by the ensuing hyper-\Ka\ space. 
In particular, when the \Ka\ space is symmetric or homogeneous, then 
the hyper-\Ka\ space is also symmetric or homogeneous, 
respectively. The material of this subsection covers some of 
the results presented in \cite{CecFerGir} and the 
relation with the work of \cite{HKLR}. Furthermore we   
discuss the behaviour under symplectic reparametrizations 
of the special hyper-K\"ahler manifold.   

The bosonic Lagrangian follows from \eqn{3dvlagr}. It can be 
rewritten as
\bea
4\pi\, \lagr &=& -N_{IJ} \Big(\pa_\m X^I \,\pa^\m \bar X^I + 
\ft14 \pa_\m A^I\,\pa^\m A^J \Big) \nonumber \\ 
&& - N^{IJ} \Big(\partial_{\mu}B_I- \ft12(F+\bar 
F)_{IK}\partial_{\mu}A^K\Big)
\Big(\partial^{\mu}B_J- \ft12(F+ 
\bar{F})_{JM}\partial^{\mu}A^M\Big) \,.\label{tormetr}
\eea
When the coordinates $A$ and $B$ are frozen to constant values,  
we have a special K\"ahler space parametrized by the 
coordinates $X^I$. Alternatively, freezing the special \Ka\ 
coordinates yields the torus $T^{2n}$. 
To describe the resulting $(4n)$-dimensional  
hyper-\Ka\ space, one must specify the metric and 
three covariantly constant complex structures, from which three
closed two-forms can be defined.

The metrics \eqn{tormetr} form a subclass of hyper-\Ka\ metrics
constructed by \cite{HKLR} using the Legendre-transform method.
The latter are characterized by the presence of at least $n$ 
abelian isometries, which are triholomorphic so that they leave 
the metric as well as the closed two-forms  
invariant. In \eqn{tormetr}, they correspond to constant shifts
in  $A^I$ and $B_I$. Hyper-\Ka\ metrics with 
at least $n$ triholomorphic abelian isometries can be written in 
the general form \cite{PedPo} 
\begin{equation}
{\rm d}s^2=U_{IJ}(x)\,{\rm d}\vec{x}^I\cdot {\rm d}\vec{x}^J+(U^{-1}(x))^{IJ}
\Big({\rm d}\varphi_I+\vec{W}_{IK}(x)\cdot {\rm 
d}\vec{x}^K\Big)\Big ({\rm d}\varphi_J+\vec{W}_{JL}(x)\cdot {\rm 
d}\vec{x}^L\Big )\ .\label{genmetr}
\end{equation}
Here, the coordinates are split according to
$\{\vec{x}^I,\varphi_I\}$. The $n$ vectors $\vec x{}^I$ 
comprise $3n$ components $\vec{x}^{I\Lambda}$, where $\Lambda=1,
2,3$; the remaining coordinates $\varphi_I$ are subject to the shift 
isometries.
The tensors $U_{IJ}$ and $\vec{W}_{IJ}$ are independent of $\varphi_I$
and satisfy the hyper-\Ka\ equations 
\begin{equation}
\partial_J^\Lambda W^\Sigma_{KI}-\partial^\Sigma_K W^\Lambda_{JI}=\varepsilon^{\Lambda
\Sigma \Pi}\partial^{\Pi}_JU_{KI}\ ,
\end{equation} 
where $\partial_I^\Lambda=\partial/\partial x^{\Lambda I}$. 
{}From this it follows that  
$\partial ^\Lambda_I U_{JK}=\partial ^\Lambda_J U_{IK}$.
The three hyper-\Ka\ two-forms, given in \cite{HKLR}, can be 
rewritten as (see e.g. \cite{2forms}) 
\begin{equation}
\omega^\Lambda=({\rm d}\varphi_I+\vec{W}_{IJ}\cdot {\rm 
d}\vec{x}^J)\wedge {\rm d}x^{\Lambda I}+U_{IJ}\, 
\varepsilon^{\Lambda \Sigma \Pi}{\rm d}x^{\Sigma I}\wedge {\rm 
d}x^{\Pi J}\ .\label{gen2forms} 
\end{equation}
Clearly, they are invariant under constant shifts of $\varphi_I$, 
so that these isometries are indeed triholomorphic.
In the case of  \eqn{tormetr}, we have coordinates 
$\vec{x}^I=(\mbox{Re}X^I,\mbox{Im}X^I,-\ft12A^I)$, 
$\varphi_I=B_I$ and 
\begin{equation}
U_{IJ}= N_{IJ}\,, \qquad \vec{W}_{IJ}=(0,0,F_{IJ}+\bar 
F_{IJ})\ .\label{sol} 
\end{equation}
For this solution both $U$ and $\vec{W}$ are 
determined by a single holomorphic function $F$,
independent of $A^I$. It can be shown that $F$ is proportional to 
the holomorphic function that appears in the contour-integral 
representation (cf. \cite{HKLR}) 
of the solution \eqn{genmetr}. Note also that $\vec{W}_{IJ}$ is 
symmetric. Other examples of hyper-\Ka\ metrics of the type 
\eqn{genmetr} are Taub-NUT  
and the asymptotic metric on the moduli space of $N$ SU(2) BPS 
monopoles. These metrics appear in the effective actions of 
three-dimensional $N=4$ SU($N$) gauge theories 
\cite{3dmirror}. They are not in the class \eqn{sol}.

To write down the hyper-\Ka\ forms and discuss symplectic transformations,
it is convenient to use the complex coordinates $Y_I$ (cf. \eqn{defY}).
In terms of the fields $X^I$ and $Y_I$ the bosonic Lagrangian reads
\bea
4\pi\,\lagr &= &- N_{IJ}\, \pa_\m X^I\pa^\m\bar X^J \\
&& -N^{IJ} \Big(\pa_\m Y_I + 
i N^{KL} (Y-\bar Y)_L \pa _\m F_{IK} \Big)\Big(\pa^\m \bar Y_J + 
i N^{MN} (Y-\bar Y)_N \pa ^\m \bar F_{JM} \Big)\,. \nonumber 
\eea
At this point we note the identity 
\be
\pa_\m Y_I + 
i N^{JK} (Y-\bar Y)_K \,\pa_\m F_{IJ}= 
(\pa_\m Y_I -\G^K_{IJ} \,\pa_\m X^J \, Y_K) -i F_{IJK}\, \pa_\m 
X^J \,N^{KL} \bar Y_L \,,\label{covY}
\ee 
where the first term is just the 
K\"ahler covariant derivative of special geometry with the 
connection given in \eqn{specialconcurv}; the second 
term is separately covariant with respect to symplectic 
reparametrizations, as can be easily verified from \eqn{sympltr}. 
Therefore the above Lagrangian is invariant 
under the symplectic reparametrizations, as we claimed already in 
the previous subsection. 

The combined $(X,Y)$ space is a hyper-K\"ahler space with \Ka\
potential \cite{CecFerGir}
\be
K(X,Y,\bar X,\bar Y)=  i X^I \,\bar F_I(\bar X) - i\bar X^I\, 
F_I(X)  - \ft12 (Y-\bar 
Y)_I\,N^{IJ}(X,\bar X)\, (Y-\bar Y)_J\,. \label{Kpotential2}
\ee
Under symplectic reparametrizations 
$K$ changes by a K\"ahler 
transformation, 
\be
\tilde K(\tilde X,\tilde Y,\tilde{\bar X},\tilde{\bar Y})= K(X,Y,
\bar X,\bar Y)   +\ft12i{\cal Z}^{IJ}(X)\,  
Y_IY_J -\ft12 i\,\bar{\cal   
Z}^{IJ}(\bar X)\,\bar Y_I\bar Y_J \,, \label{symplK}
\ee
where $\tilde K$ is evaluated on the basis of the new function 
$\tilde F$ and ${\cal Z}(X)$ is the symmetric holomorphic tensor 
defined in  
\eqn{symplv}. This does not imply that the \Ka\ metric takes the 
form of a symplectically covariant tensor, because the coordinates 
$Y_I$, unlike the special \Ka\ coordinates 
$X^I$, do not transform as coordinates but as symplectic 
vectors.\footnote{%
  A similar situation is present in Calabi's construction of hyper-K\"ahler spaces 
  on cotangent bundles with coordinates $(X^I,Y_I)$ \cite{Calabi}.
  The corresponding K\"ahler potential is $K=i(X^I\bar F_I-\bar X^IF_I)
  -Y_IN^{IJ}\bar Y_J$, and is invariant under symplectic transformations.
  An essential difference, however, is that Calabi's metric does not
  possess the  same (triholomorphic) isometries as  the metric 
  described above.} %
To see this, one first computes the metric from the derivatives of the \Ka\ potential.
In the coordinates $z^a = (X^I, Y_J)$ we find
\be
g_{a\bar b} = \pmatrix{ (N+ P\,N^{-1}\bar P)_{IK} & (P \,
N^{-1})_I{}^L\cr
\noalign{\vskip 4mm}
(N^{-1} \bar P)^J{}_K & (N^{-1})^{JL} \cr}\,,
\ee
where we have defined the symmetric tensor $P_{IJ}=i F_{IJK} N^{KL}
(Y-\bar Y)_L$. Under a symplectic reparametrization, $P_{IJ}$ transforms as
\be
P_{IJ} \to \Big[P_{KL} + F_{KLM} \,{\cal Z}^{MN}Y_N\Big]\,   
[{\cal S}^{-1}]^K{}_{I}\,[{\cal S}^{-1}]^L{}_{J} \,,
\ee
which implies that the metric is not symplectically covariant.

The inverse metric satisfies the relation
\be
\O_{ac}\,g^{c\bar d}\,\O_{\bar d\bar b} = - g_{a\bar b}
\,,\label{herm}
\ee
where $\O_{ab}$ is a covariantly constant antisymmetric tensor, 
\be
\O_{ab} = \pmatrix{ 0& \d_I{}^L\cr
\noalign{\vskip 1mm}
- \d^J{\!}_K&0 \cr}\,. 
\ee
The covariant constancy follows from \eqn{herm}. As a result $\O$ 
commutes with the holonomy group. Complex structures are then defined by
\be
J^3= \pmatrix{-i \d^a{}_b& 0\cr
    0&i \d^{\bar a}{}_{\bar b}\cr}\,, \qquad J^\a = 
\pmatrix{0& \alpha\O_{a}{}^{\bar b}\cr
    \bar \alpha\O_{\bar a}{}^{b}&0\cr} \,, 
\label{complexstructures}
\ee
with $\alpha$ a phase factor and $\O_a{}^{\bar b}=\O_{ac}g^{c\bar b}$. 
Choosing $\a=1,-i$ corresponding to, 
respectively, $J^2$, $J^1$, the 
matrices satisfy 
\be
J^\Lambda\,J^\Sigma = -{\bf 1}\,\d^{\Lambda\Sigma} -
\varepsilon^{\Lambda\Sigma\Pi}\,J^\Pi\,. 
\ee
Finally, the hyper-\Ka\ forms can be computed
from \eqn{gen2forms} or, equivalently, from the complex structures.
One finds
\bea
\o^3 &=& -i K_{X\bar X} \,{\rm d}X\wedge  {\rm d}\bar X -i 
K_{X\bar Y} \,{\rm d}X\wedge{\rm d}\bar Y- i K_{Y\bar X} \,{\rm 
d}Y\wedge {\rm d}\bar X - i K_{Y\bar Y}\, {\rm d}Y \wedge {\rm 
d}\bar Y  \,,\nonumber \\
\o^+&=&  {\rm d}X^I \wedge {\rm d}Y_I \,,\qquad
\o^-= {\rm d}\bar X^I \wedge {\rm d}\bar Y_I \,.
\label{cl2form}
\eea
Observe that $\o^{\pm}$ is purely (anti-)holomorphic, as already mentioned
in \cite{Calabi,HKLR}. This will be important when we discuss the central
charges in section~5.
These two forms are closed 
so that locally they can  
be written as an exterior derivative of the following one-forms,
\bea
A^3 &=& \ft12 i K_X\, {\rm d}X + \ft12 i K_Y \,{\rm d}Y- \ft12 i 
K_{\bar X} \,{\rm d}\bar X -\ft12 i K_{\bar Y}\, {\rm d}\bar Y
\,,\nonumber \\
A^+ &=& \ft12X^I \,{\rm d}Y_I - \ft12Y_I \,{\rm d}X^I \,,\qquad A^-= 
\ft12\bar X^I \,{\rm d}\bar Y_I - \ft12\bar Y_I \,{\rm d}\bar 
X^I\, . 
\eea 
Under symplectic reparametrizations, these one-forms are 
invariant up to an exact form. For $A^3$ this follows from 
\eqn{symplK} and for $A^\pm$ this can be seen from noting that 
the second term is manifestly symplectically invariant, whereas 
the first term equals the second up to an exact form. Therefore 
the corresponding hyper-\Ka\ two-forms are symplectically 
invariant. For $\o^\pm$ this 
can also be seen directly by observing that  
replacing the one-forms ${\rm d}Y$ by the symplectically 
covariant forms  ${\rm d}Y_I + i N^{JK} Y_K {\rm d} F_{IJ}$,  
does not change $\o^\pm$.  
Note, however, that the corresponding tensors $J^\L$ are {\it not} 
symplectically covariant, just as the metric was not a covariant 
tensor. This is again related to the  
fact that the symplectic reparametrizations act differently on 
the $X^I$ and $Y_I$. 

\subsection{Isometries}
As explained in detail in \cite{ssss} the isometry group of a 
special \Ka\ manifold extends in a characteristic way when 
performing the \cmap. The additional isometries are called {\it 
extra} symmetries when their origin can be understood directly from the 
four-dimensional gauge transformations, or {\it hidden} symmetries 
when their existence is not generic and depends on special 
properties of the manifold. In \cite{ssss} this was discussed for 
special quaternionic manifolds (i.e., in the case of local 
supersymmetry). In this subsection, we give a similar discussion 
for the special hyper-\Ka\ manifolds. Here the {\it extra} 
symmetries follow directly from the gauge  
symmetry in four dimensions and correspond to constant 
shifts in $A^I$ and in $B_I$, as we discussed previously.  
In the complex basis the extra isometries take the form
\be
\delta Y_I=\beta_I- F_{IJ}\alpha^J\ , \label{shift}
\ee
with real parameters. They leave the metric invariant and also the 
closed two-forms $\omega ^\Lambda$ of \eqn{cl2form}. By 
definition, such isometries are called triholomorphic.
    
Apart from these, there can be isometries corresponding to 
duality symmetries of the  
original four-dimensional action of the vector multiplets. These 
isometries are associated with the symplectic reparametrizations, 
leaving the function $F$ unchanged. Because the two-forms $\o^\L$ 
are symplectically invariant, these isometries are also triholomorphic.
There can be additional isometries of the special \Ka\ 
manifold that do not leave the full action invariant \cite{ssss,
sssl}. Those isometries do not take the form of symplectic 
reparametrizations and will in principle not correspond to 
isometries of the hyper-\Ka\ manifold.  

Just as for the special 
quaternionic manifolds, we find that {\it hidden} 
symmetries for the hyper-\Ka\ manifolds are subject to certain 
nontrivial conditions. But unlike the quaternionic case, 
the conditions seem impossible to satisfy unless one makes a rather 
simple choice for the function $F(X)$.  
Before proceeding to derive the conditions for general isometries, 
we make the following observation. Obviously the 
commutator of an infinitesimal isometry and a supersymmetry 
variation defines a fermionic symmetry. However, we know that the fields 
$X^I$ and $Y_I$ transform only under supersymmetries with 
positive-chirality parameters.  Unless the isometries are 
holomorphic, we will thus generate new supersymmetries of the wrong 
chirality. These can not be accomodated by the standard 
supersymmetry algebra and the theory can only be invariant under 
them if it contains noninteracting sectors, i.e., if the 
model is reducible and the target space is a local product space 
(this argument is identical to the one used in \cite{DWVN} for 
two-dimensional sigma models with torsion). So without loss of 
generality, we may assume that the isometries are holomorphic. 

With this in mind we first study the variations of the action under 
an arbitrary infinitesimal isometry that are quadratic in the 
derivatives of the fields $A^I$ and $B_I$. This leads to the result that the 
variation of $F_{IJ}$ must take the form
\be
\d F_{IJ} =  N_{IK}\,N_{JL} \, {\pa^2 f\over \pa\bar Y_K \pa\bar 
Y_L}  \,, \label{delF}
\ee
where $f$ is some real function of $Y,\bar Y,X, \bar X$. 
Furthermore the transformation rule for $Y_I$ can be written as   
\be
\d Y_I= -i N_{IJ}\, {\pa \over \pa\bar Y_J} \bigg[ 2f + (Y_K- \bar 
Y_K){\pa f\over \pa\bar Y_K}\bigg] + i N_{IJ}\L^{JK}\bar Y_K\,,
\label{delY}  
\ee
where the quantity $\L^{IJ}(X,\bar X)$ is independent of $Y$ and 
$\bar Y$ and 
antisymmetric in $I,J$ so that it cannot be incorporated into 
the first term for $\d Y_I$. 
{}From the fact that the right-hand side of \eqn{delF} must be 
independent of $\bar Y$, it follows that the function $f$ depends 
at most quadratically on $\bar Y$. (Obviously the same conclusion 
can be drawn for the $Y$-dependence.) Therefore the first  
term in \eqn{delY} is $\bar Y$-independent. Because $\d Y$ itself 
must also be independent of $\bar Y$, it follows that $\L^{IJ}=0$.   

Subsequently, consider the mixed variations in the Lagrangian, 
proportional to a derivative of $A$ or $B$ and $\bar X$. This 
leads to conditions for the derivatives of $\d X^I$ with respect 
to $A^I$ and $B_I$, which can be integrated. Specifically, we 
find two restrictions,
\be
\d X^I \pm i N^{IJ} {\pa f\over \pa \bar X^J}\bigg\vert_{A,B} = 
\ft12 P^I_\pm \,,
\ee
where $P_+^I$ depends on $X,\bar X, \bar Y$ and $P_-^I$ depends 
on $X,\bar X, Y$. The holomorphy of $\d X^I$ implies that 
$(P_++P_-)^I$ depends  
only on $X$ and $Y$. Therefore it follows that $N^{IJ}\pa f/\pa \bar 
X^J\vert_{A,B}$ must be independent of $\bar Y$ and $\bar X$, up 
to terms that depend exclusively on $X,\bar X$. 
The holomorphy in $Y$ restricts $f$ to the following form, 
\bea
f(X,\bar X, Y,\bar Y) &=& [N(Y-\bar Y)]^I [N(Y-\bar Y)]^J 
O_{IJ}(X,Y)  \nonumber \\ 
&& +i  [N(Y-\bar Y)]^I \L_I(X,Y) + \tilde f(X,\bar X, Y)\,. 
\label{ansatz}
\eea
The holomorphic functions $O_{IJ}$ and $\L_I$ can now be expanded 
in powers of $Y$. Note that the first one is at most quadratic 
and the second one at most cubic in $Y$. Also the nonholomorphic 
function $\tilde f$ can be expanded in $Y$, up to fourth order. 

The reality of $f$ yields a large number of restrictions. For 
instance, the $Y$-expansion coefficients of $\tilde f$ and $\L$  
are related,
\be
[\L_I(X) -\bar \L_I(\bar X)]^{J_1\cdots J_n} = i n N_{IJ}\,\tilde 
f^{JJ_1\cdots J_n}(X,\bar X)\,,
\ee
where the $\tilde f^{IJK\cdots}$ must be real. 
On the other hand, holomorphy in $X$ restricts the $\tilde 
f^{IJK\cdots}$ to the form
\be
\tilde f^{KL\cdots} (X,\bar X) = (\bar X^IF_{IJ} -\bar F_J) \, 
g^{J,KL\cdots} (X) + h^{KL\cdots}(X)\,.
\ee
Combining these constraints seems to lead inevitably to the 
conclusion, at least for nontrivial functions $F(X)$, that the 
$\tilde f^{IJ\cdots}$ must be constant.  
In that case we may rewrite \eqn{ansatz} in terms of $A$ and $B$, 
and observe that there is no $\bar X$-dependence anymore. 
However, the function $f$ must be real, so that we conclude that 
it can be written as the sum of a 
function of $A$ and $B$ and a function of $X$ and $\bar X$. 
The latter function can be ignored.
The independence of $X$ and $\bar X$ now implies that 
\eqn{ansatz} is a real polynomial in $A$ and 
$B$ that is at most of order two. The terms linear in $A$ and $B$ 
characterize the shift symmetries \eqn{shift}, whereas the quadratic terms 
correspond to the isometries embedded in the symplectic 
reparametrizations of the special \Ka\ space. The latter can be 
verified by showing that \eqn{delF} and \eqn{delY} take the form of an 
infinitesimal symplectic reparametrization as follows from the 
first equation of \eqn{sympltr} and \eqn{Ytrans}, respectively. 


\section{Hypermultiplets}
\setcounter{equation}{0}
Hyper-\Ka\ spaces serve as target spaces for hypermultiplets. One 
of our goals is to understand the relation between special \Ka\ 
and special hyper-\Ka\ at the level of the full actions for vector 
multiplets and hypermultiplets, including the fermions. Before 
doing this we briefly review the derivation of the Lagrangian for  
hypermultiplets in four spacetime dimensions. Our analysis, which 
is self-contained,  
is closely related to the one presented in  
\cite{BagWit}. However, our results are cast in a somewhat different 
form in order to facilitate the comparison with the models that 
emerge from the vector multiplets under the action of the \cmap. 
Furthermore, we find that a certain restriction found in 
\cite{BagWit} is unnecessary and in fact too restrictive. 

We assume $4n$ real scalars $\phi^A$ and $2n$ positive-chirality 
spinors $\zeta^{\bar \a}$ and $2n$ negative-chirality spinors 
$\zeta^\a$, which are related by conjugation (so that we have 
$2n$ Majorana spinors). Therefore, under complex conjugation 
indices are converted according to 
$\a\leftrightarrow \bar \a$, while, just as before,  SU(2) 
indices $i,j, \ldots$ are raised and lowered. The 
supersymmetry transformations are parametrized in terms of 
certain $\phi$-dependent quantities $\g^A$ and $V_A$ as
\bea
\d\phi^A&=& 2\Big(\g^A_{i\bar\a} \,\bar\e^i \zeta^{\bar \a} + 
\bar\g^{Ai}_{\a} \,\bar\e_i \zeta^\a\Big)\,,\nonumber\\
\d\zeta^{\bar \a}&=& \bar V_A^{i\bar\a} \,\pa\slash\phi^A\e_i 
-\d\phi^A\,\bar\G_{A}{}^{\!\bar\a}{}_{\bar \b} \,\zeta^{\bar \b} \,,
\nonumber\\ 
\d\zeta^\a  &=&  V_{A\,i}^\a \,\pa\slash\phi^A\e^i 
-\d\phi^A\, \G_{A}{}^{\!\a}{}_\b\,\zeta^\b \,. \label{4dhsusy}
\eea
The definition of $\G$ and $\bar \G$ will be discussed shortly. 
As it turns out, with the proper definition, the above ansatz  
comprises the full supersymmetry transformation laws. Observe 
that the variations are consistent with a U(1) chiral invariance 
under which the scalars remain invariant, which we will denote by 
${\rm U(1)}_{\rm R}$ to indicate that it is a subgroup  
of the automorphism group of the supersymmetry algebra. However, 
for generic $\g^A$ and $V_A$, the SU(2)$_{\rm R}$ part of the 
automorphism group cannot be realized consistently. In the 
above, we only used that $\z^\a$ and $\z^{\bar \a}$ are related 
by complex conjugation. Our notation is similar 
but not identical to the one used in \cite{DWLVP}. 

A first condition on the quantities $\g^A$ and $V_A$ follows from 
the closure of the supersymmetry transformations \eqn{4dhsusy} on 
the scalars. This yields the Clifford-like condition
\be
\g^A_{i\bar\a} \,\bar V_B^{j\bar\a} + \bar \g^{A\,j}_\a \,V_{B\,
i}^\a  = \d^j_i\,\d^A_B\,. \label{clifford}
\ee

Subsequently let us turn to the action, which we parametrize 
as 
\be
4\pi\, \lagr= -\ft12 g_{AB}\,\pa_\m\phi^A\pa^\m\phi^B  
-G_{\bar \a \b}\Big( \bar\zeta^{\bar \a} D\!\slash \,\zeta^\b +  
\bar\zeta^\b D\!\slash \,\zeta^{\bar\a}\Big) + \lagr(\zeta^4) 
\,, \label{4dhlagr1}
\ee
where $G_{\bar\a \b}$ is a hermitean metric\footnote{%
   A possible antihermitean part can be  absorbed into the 
   Noether term, modulo a total derivative. In principle, it is 
   possble to absorb $G$ into the definition of the fermion fields, 
   but we refrain from doing so for reasons that will become clear 
   in due course.}, %
and we use the covariant derivatives
\be
D_\m \zeta^\a= \pa_\m \zeta^\a + \pa_\m\phi^A\, 
\G_{A}{}^{\!\a}{}_{\!\b} \,\zeta^\b\,, 
\quad
D_\m \zeta^{\bar\a}= \pa_\m \zeta^{\bar \a}  +\pa_\m\phi^A\,\bar 
\G_{A}{}^{\!\bar\a}{}_{\bar \b} \,\zeta^{\bar \b} \,.
\ee
The Noether term thus takes the following form,
\be
4 \pi\, \lagr_{\rm N}= \Big[ \bar \G_{A}{}^{\!\bar \g}{}_{\bar\a}\, G_{\bar \g\b}- 
G_{\bar \a\g}\,\G_{A}{}^{\!\g}{}_{\b} \Big]\,   \bar \zeta^{\bar 
\a}\pa\slash\phi^A\zeta^\b \,. 
\ee
Observe that only a linear combination of the two connections 
appears in the action. 

Considering various terms of the supersymmetry variation of the action 
\eqn{4dhlagr1} leads to further conditions. 
Cancellation of the variations proportional to $\pa^2\phi^A$ 
implies
\be
g_{AB}\, \g^B_{i\bar \a} = G_{\bar\a\b}\, V_{A\,i}^\b \,,\qquad
g_{AB}\, \bar\g^{B\,i}_\a = G_{\bar\b\a}\, \bar V_A^{i\bar\b} 
\,.   \label{gammaV}
\ee
Then variations proportional to $\pa_\m\phi^B\,\pa_\n\phi^C$ require 
\bea
2 G_{\bar\b\a}\, D_B V_{A\,i}^\a   + D_B G_{\bar\b\a}\, 
V_{A\, i}^\a  &=&0 \,, \nonumber \\
2 G_{\bar\b\a}\, D_B \bar V_A^{i\bar \b}   + D_B 
G_{\bar\b\a}\, \bar V_A^{i\bar \b}  &=&0\,.   
\label{var1}
\eea
Note that the first covariant derivative in \eqn{var1} contains 
also the Christoffel symbol $\{A;BC\}$. Now redefine the connections 
according to 
\bea
G_{\bar\b\g} \G_{A}{}^{\!\g}{}_{\a} +  \ft12 D_A G_{\bar\b\a} &\to& 
G_{\bar\b\g}  \hat \G_{A}{}^{\!\g}{}_{\a}  \,, \nonumber \\
G_{\bar\g\a} \bar \G_{A}{}^{\!\bar\g}{}_{\bar \b} + \ft12 D_A 
G_{\bar\b\a}  &\to& G_{\bar\g\a}  \hat{\bar 
\G}_{A}{}^{\!\bar\g}{}_{\bar\b}\,.   
\eea
Taking the difference, one sees that this modification does not 
modify the Noether term. Furthermore, one can verify that 
the $\g^A$ tensors are covariantly constant with respect to the 
connection $\hat \G$, and so is the metric $G_{\bar\a\b}$. Thus we replace 
the connections everywhere by the new connection and drop the 
caret. These are then the connections that appear in the 
variations of the spinor fields in \eqn{4dhsusy} and, as it turns 
out, no additional terms quadratic in the spinor fields are 
required in these transformation rules.

According to the above results we define four real, antisymmetric 
covariantly constant tensors, 
\be
J^\L_{AB} = i \g_{Ai\bar \a}\, \bar V_B^{ j\bar \a}\, 
(\s^\L)^i{}_j \,,\qquad  (\L=1,2,3) \label{cstructures}
\ee
and
\be
C_{AB} = i(\g_{Ai\bar \a}\, \bar V_B^{ i\bar \a} - g_{AB})\,. 
\label{defC}
\ee
It follows that $C$ must vanish, so that $\g$ and $\bar V$ 
are each others inverse, 
\be
\bar V^{i\bar \a}_A \, \g^A_{j\bar \b} = \d^i_j\, 
\d^{\bar \a}_{\,\bar \b}\,. \label{inverse}
\ee
The precise analysis leading to this result is somewhat subtle, and 
is based on an extension of the arguments used in \cite{AGF}. 
It makes use of the fact that the five covariantly constant 
two-rank tensors, the metric, the $J^\L$ and $C$, 
and products thereof,  must commute with the 
curvature tensor and therefore with the holonomy group. The 
latter can act reducibly, so that the target space 
factorizes and the model decomposes into 
the sum of several independent models. If the holonomy group acts 
irreducibly, then according to Schur's lemma, the algebra 
generated by the above tensors must be a division algebra. This 
implies a degeneracy between the tensors \eqn{cstructures} and 
\eqn{defC}. Combining this fact  with the Clifford property leads 
to \eqn{inverse}. 
 
{}From \eqn{inverse} it then follows directly  that the $J^\L$ are 
complex structures, satisfying 
\be
J^\L\,J^\S = - {\bf 1} \d^{\L\S} - \varepsilon^{\L\S\Pi} J^\Pi\,,
\ee
reflecting the well-known result that the target space must be 
hyper-\Ka. 

Furthermore we note the existence of covariantly constant 
antisymmetric tensors,  
\be
\O_{\bar\a\bar\b} =\ft12  \varepsilon^{ij}\,g_{AB}\, \g^A_{i\bar 
\a}\,\g^B_{j\bar \b}\,,
\quad 
\bar\O^{\bar\a\bar \b} =\ft12  \varepsilon_{ij}\,g^{AB}\, \bar 
V_A^{i\bar \a}\,\bar V_B^{j\bar \b}\,,
\ee
satisfying
\be
\varepsilon_{ij} \, \O_{\bar \a\bar \b} \, \bar V_A^{j\bar \b} = 
g_{AB} \,\g^B_{i\bar \a}\,. \label{psreal}
\ee
According to \eqn{gammaV} and \eqn{psreal} the $\g$ and $V$ tensors 
are linearly related and pseudo-real. Therefore the tensor $\O$ 
is also pseudo-real and it satisfies  
\be
\O_{\bar\a\bar\g}\,\bar \O^{\bar\g\bar\b} = -\d^{\bar \b}_{\bar 
\a}\,.
\ee

The existence of covariantly constant tensors implies a variety 
of integrability conditions for the curvature tensors. {}From the 
constancy of $G_{\bar\a\b}$ and $\O_{\bar\a\bar \b}$ we obtain,
\be
R_{AB}{}^{\!\bar\b}{}_{\bar\a} = - G_{\bar\a\g }\, G^{\d\bar\b} \,
R_{AB}{}^{\!\g}{}_\d \,, \qquad 
R_{AB}{}^{\!\bar \g}{}_{[\bar\a}\,\O_{\bar \b]\bar\g} =0\,.
\ee
These conditions imply that $R_{AB}{}^{\!\a}{}_\b$ 
takes values in $sp(n)\cong usp(2n;{\bf C})$ so that the holonomy 
group acts symplectically on the fermions. 

Furthermore, constancy of the $\g$ tensor implies
\be
R_{ABD}{}^C \,\g^D_{i\bar\a} - R_{AB}{}^{\!\bar \g}{}_{\bar \a} \,
\g^C_{i \bar\g} = 0\,. \label{covconst}
\ee
{}From this result one proves that Riemann curvature and the 
Sp$(n)$ curvature are related,  
\bea 
&& R_{AB}{}^{\,\bar \b}  {}_{\bar\a}= \ft12 R_{ABE}{}^C\,\g^E_{i\bar 
\a}\,\bar V^{i\bar \b}_C\,, \nonumber \\
&& R_{ABD}{}^C = R_{AB}{}^{\!\bar \b}{}_{\bar\a}\,
\g^C_{i \bar\b}\, \bar V_D^{i\bar \a} \,.
\eea
Using the pair-exchange property of the Riemann tensor and 
contracting with $\g^C\, \bar \g^D$ one derives
\be
R_{AB}{}^{\!\bar \b}{}_{\bar \a}   = \ft12 W_{\bar \a \e \bar \g\d} \, 
\bar V^{i\bar \g} _A\, V_{Bi}^\d\, G^{\e\bar \b}  \,, 
\label{RABabW}
\ee
where  
\be
W_{\bar \a \b \bar \g\d} = 
R_{AB}{}^{\!\bar\e}{}_{\bar \g} \,\g^A_{i\bar\a}\,\bar \g^{iB}_\b\, 
 G_{\bar \e\d}  = \ft12 R_{ABCD} \,\g^A_{i\bar\a}\,\bar \g^{iB}_\b\, 
\g^C_{j\bar\g} \,\bar \g^{jD}_\d\, . \label{defW}
\ee
The tensor $W$ can be written as $W_{\a\b\g\d}$ by contracting 
with the metric $G$ and the antisymmetric tensor $\O$. It then 
follows that $W_{\a\b\g\d}$ is symmetric in symmetric index pairs 
$(\a\b)$ and $(\g\d)$. Using the Bianchi identity for Riemann 
curvature, which implies $g_{D[A}R_{BC]}{}^{\!\bar \b}{}_{\bar\a} 
 \,\g^D_{i\bar\b} = 0$, one shows that it is in fact symmetric in 
all four indices. 

Hence all the curvatures are expressed in terms of the fully 
symmetric tensor $W_{\a\b\g\d}$. {}From this result many other 
identities for the curvatures can be derived. In particular we 
note the identity   
\be
R_{AB}{}^{\!\bar\g}{}_{[\bar\a} \,\g^B_{\bar\b] i} = 0\,, 
\label{crucialid}
\ee
which plays a crucial role in proving the supersymmetry of the 
action. For that, one needs to include a four-fermion interaction 
into the Lagrangian, equal to 
\be
4\pi\,\lagr(\zeta^4) = -\ft14  W_{\bar \a\b\bar\g\d}\, \bar \zeta^{\bar \a} 
\g_\m\zeta^{\b}\,\bar \zeta^{\bar \g} \g^\m\zeta^\d \,.
\ee

All the above results are closely related to the ones derived 
long ago in \cite{BagWit}. One feature that is different is the  
presence of a fermionic metric, which, as we will demonstrate in 
the next section,  is important in exhibiting the effect of symplectic 
reparametrizations for models in the image of the \cmap. Another 
feature concerns the condition imposed in \cite{BagWit} that
$\g^B_{i\bar\a} \,\bar \g^{C\,i}_\b + 
\g^C_{i\bar\a} \,\bar \g^{B\,i}_\b$ be proportional to
the product of $g^{BC}$ and $G_{\bar \a\b}$ and 
inversely proportional to the number of hypermultiplets $n$. 
We found no need for this condition. 
In fact, it is in contradiction with the case of 
free fields, where no $1/n$ terms can arise.

\section{Applying the mirror map}
\setcounter{equation}{0}
{}From the material of the previous sections we will explicitly 
extract the basic quantities of the hyper-\Ka\ space that emerges 
from the four-dimensional $N=2$ vector multiplets under the action 
of the \cmap. Before doing so, it is important that we first 
discuss the extension of the chiral ${\rm SU}(2)_{\rm R}\times {\rm 
U}(1)_{\rm R}$ automorphism group of the supersymmetry algebra in 
four spacetime dimensions to SO(4). Of course, it is well known 
that the automorphism group in three dimensions contains SO(4), but 
we are interested in the way this extension is realized, namely 
by promoting the ${\rm U}(1)_{\rm R}$ group to SU(2). 
With the aforementioned ${\rm SU}(2)_{\rm R}$ one thus obtains 
the group $({\rm SU}(2)\times {\rm SU}(2))/{\bf Z}_2\cong {\rm SO(4)}$. 

In section~3 we already made reference to the fact that the 
independent combinations of four-dimensional gamma matrices that 
commute with  
the three-dimensional ones, constitute an $su(2)$ algebra. 
Therefore spinors $\epsilon^i$ in a four-dimensional 
spacetime, which transform under a chiral ${\rm SU}(2)_{\rm 
R}\times {\rm U}(1)_{\rm R}$ group,  can in principle transform 
under a bigger group after descending to 
three dimensions. However, we are not interested in any such 
extension, but only in those that constitute a subgroup of 
the automorphism group of the supersymmetry algebra in three 
spacetime dimensions.  

To understand the fate of the $su(2)$ let us momentarily consider 
$N=1$ supersymmetry in four spacetime dimensions. The four-dimensional 
automorphism group contains a chiral U(1). According to the above 
arguments this group can be extended to SU(2) in the reduction to 
three spacetime dimensions; its generators are just 
proportional to the three hermitean matrices $\hat \s^1, \hat\s^2, 
\hat \s^3$ that were defined in section~3. This 
SU(2) group is consistent with the supersymmetry algebra, but it  
cannot be realized on Majorana spinors. The Majorana constraint 
requires the  phases appropriate to the group SL(2), 
which, in turn, is not consistent with the supersymmetry 
algebra. So, unless one doubles the spinors, the automorphism 
group U(1) remains unextended when descending to three spacetime 
dimensions.  

Starting from $N=2$ in four dimensions, on the other hand, 
naturally incorporates such a doubling of spinors. The spinor 
doublets then transform under chiral ${\rm SU}(2)_{\rm R}\times 
{\rm U}(1)_{\rm R}$ and the extension of the U(1) group to 
SU(2) is automatic. Starting with a (nonchiral) Majorana 
doublet $\e^i$ (which comprises eight real independent 
components), the ${\rm SU}(2)_{\rm R}$ transformations act 
according to
\be 
\e^{i} \to  \bigg[ U^i{}_j\,\Big({{\bf 1}+\hat\s^2\over 2}\Big) + 
\overline{U^i{}_j} \,\Big({{\bf 1}-\hat\s^2\over 2}\Big) \bigg] \e^{j}\,.
\ee
In three spacetime dimensions the group ${\rm U}(1)_{\rm R}$ is 
extended to SU(2), with matrices $\hat U$ that are asssociated with the  
generators $\hat\s^a$. They act according to
\be
\e^{i} \to \bigg[ \hat U  \,\Big({{\bf 1}+ \s^2\over 2}\Big)^i{}_j + 
\bar{\hat U }\, \Big({{\bf 1}-\s^2\over 2}\Big)^i{}_j  \bigg]\e^{j}\,,
\ee
where $(\s^2)^i{}_j$ equals the skew-symmetric imaginary $\s$-matrix. 
This extra SU(2) commutes with ${\rm SU}(2)_{\rm R}$ by virtue of 
the fact that they both have a skew-symmetric invariant  
tensor, $(\s^2)^i{}_j$ and $\hat\s^2= \g^5$, satisfying $\bar 
U=\s^2 U\,\s^2$ and likewise for $\hat U$ and $\hat\s^2$. It is 
convenient to write the above transformations in infinitesimal 
form employing chiral spinor components. Defining $\hat U \approx 
{\bf 1}+ \ft12 i \hat\a_a\,\hat \s^a$, one obtains
\bea
\d\e^i &=&\ft12 i\hat \a_2\, \e^i +\ft12\varepsilon^{ij} (\hat 
\a_1+i\hat \a_3)\g^3 \e_j\,,\nonumber\\
\d\e_i &=& -\ft12i\hat \a_2\, \e_i +\ft12 \varepsilon_{ij} (\hat 
\a_1-i\hat\a_3)\g^3 \e^j\,.
\eea
The above results show that a proper basis for the extra SU(2) 
transformations is obtained by choosing
\bea
\e^+=\ft12\sqrt{2}\g^3(\e_1-i\e_2)\,,\qquad
\e^- =\ft12\sqrt{2}(\e^1-i\e^2)\,, \nonumber \\
\e_+ =\ft12\sqrt{2} 
\g^3(\e^1+i\e^2)\,,\qquad
\e_- =\ft12\sqrt{2}(\e_1+i\e_2)\,.  \label{newpara}
\eea
These spinors are eigenstates under both $\s^2$ and $\hat\s^2$ and 
transform under phase transformations under both U$(1)_{\rm R}$ 
as the SO(2) subgroup of SU$(2)_{\rm R}$. Upper- and  
lower-index spinors are related by conjugation.

Now let us consider the reduction to three dimensions of the 
actions presented in sections~2 and 4 for vector multiplets and 
hypermultiplets. As pointed out previously, the vector multiplet 
Lagrangian and supersymmetry transformations are manifestly 
covariant with respect to the ${\rm SU}(2)_{\rm R}$ group, but 
not to the group ${\rm U}(1)_{\rm R}$ (at least, not in the general 
case). Consequently, when descending to three dimensions, the 
symmetry group is not enhanced and we are left with the ${\rm 
SU}(2)_{\rm R}$ transformations and the symplectic 
reparametrizations. On the other hand, the hypermultiplet 
Lagrangian and supersymmetry transformations are generically only 
covariant with respect to the group ${\rm U}(1)_{\rm R}$ and when 
descending to three dimensions, this group is enhanced to a full SU(2)
group, with elements $\hat U$. However, consistency requires that 
this extra SU(2)  
group commutes with the holonomy group and therefore its action 
incorporates the antisymmetric tensor $\O_{\bar \a \bar \b}$ 
constructed in the previous section. Infinitesimally the SU(2) 
transformations act on the hypermultiplet fermions according to 
\bea
\d\zeta^\a &=& \ft12 i\hat \a_2\, \zeta^\a 
-\ft12 G^{\a\bar\g}\O_{\bar\g\bar\b} \,(\hat \a_1+i\hat\a_3)\g^3 
\zeta^{\bar\b}\,,\nonumber\\ 
\d\zeta^{\bar\a} &=& -\ft12 i\hat \a_2\, \zeta^{\bar\a} 
- \ft12 \bar\O^{\bar\a\bar\g}G_{\bar\g\b} \,(\hat \a_1-i\hat 
\a_3) \g^3 \zeta^\b \,. 
\eea
In other words, when systems based on both vector multiplets and 
hypermultiplets are  reduced to a three-dimensional spacetime, 
the target space factorizes into two hyper-\Ka\ manifolds which 
will both possess  
an independent SU(2) invariance group, corresponding to different  
factors of the SO(4) automorphism group of the supersymmetry 
algebra. This reflects the general 
situation in $N=4$ supersymmetric sigma models in three 
dimensions, even when coupled to supergravity. In the latter case 
the sigma model target space factorizes into 
two quaternionic spaces, whose Sp(1) holonomy groups consitute 
the two different factors of the SO(4) group \cite{dWTNic}.
This situation is typical for the case of $N=4$ supersymmetry.
 
The above observations are essential to reconcile the fermionic 
supersymmetry 
transformations \eqn{3dvsusy} with those of the hypermultiplet 
\eqn{4dhsusy}, after dimensional reduction.  
The SO(2) subgroup of SU$(2)_{\rm R}$ will play the role of 
U$(1)_{\rm R}$  after applying the mirror map and returning to four 
spacetime dimensions. Consequently, we must identify the fields 
$\zeta^{\a}$ and $\zeta^{\bar \a}$ with combinations of the vector 
multiplet spinor fields, $\O^I_i$ and $\O^{iI}$, that transform 
as eigenspinors under the SO(2) group with the proper phase 
transformations. For the spinor parameters, this means that we 
must convert to the previously introduced spinor  
parameters $\e^\pm$ and $\e_\pm$ (cf. \ref{newpara}). These 
requirements motivate us to make the following identification,
\bea
\zeta^\a &=& \Big(-\ft12\sqrt{2}\g^3(\O_1^I -i\O_2^I), 
\ft12\sqrt{2}(\O^{1I} - 
i\O^{2I})\Big)\,,\nonumber \\
\zeta^{\bar\a}  &=& \Big( -\ft12\sqrt{2}\g^3(\O^{1I} + i\O^{2I}),
\ft12\sqrt{2} (\O_1^I +i\O_2^I) \Big)\,, \label{ident}
\eea
where the relation between $\zeta^\a$ and $\zeta^{\bar\a}$ 
proceeds via Dirac conjugation and the Majorana condition.  

Let us first comment on the various factors in \eqn{ident}. As 
explained above, the identification is such that the $\zeta^{\a}$ 
transform under the SO(2) subgroup of  
SU$(2)_{\rm R}$ with a uniform phase. The $\zeta^{\bar\a}$ then 
transform with the opposite phase.  The relative factors $\g^3$ 
follow from the requirement that the fermions on the right-hand 
side, whose supersymmetry transformations follow from 
\eqn{3dvsusy}, will take a form similar to the transformations  
of the hypermultiplet fermions, as given in \eqn{4dhsusy}, 
when descending to three dimensions. Both the overall and 
relative factors of 
$\g^3$ are required to match the chirality of both sides of the 
equations. The phase factors adopted for the 
various components in \eqn{ident}, are somewhat arbitrary. They 
can be changed a posteriori by performing certain redefinitions. 
The same comment applies to the phase factors adopted in the 
definitions of the spinors \eqn{newpara}. 

In three dimensions, \eqn{ident} and \eqn{newpara} represent 
simply a different basis for the spinors that play a role in the 
vector multiplet. However, from the point of view of the 
four-dimensional Lorentz group, this choice of basis has 
nontrivial implications. 
When assuming that the newly defined spinor fields transform in 
the conventional  
way under the four-dimensional Lorentz transformations, one 
implicitly exchanges the SU$(2)_{\rm R}$ and the extra SU(2) group that 
contains U$(1)_{\rm R}$. More precisely, taking the vector 
multiplet to three dimensions, the four-dimensional gamma 
matrices are related to the three-dimensional ones, properly 
combined with the SU(2) generators denoted by $\hat \s^a$. 
Returning to four dimensions in the same way as before, but on 
the basis of the newly defined spinors, implies that the 
four-dimensional gamma matrices are  now formed from the 
three-dimensional gamma matrices combined with the SU$(2)_{\rm 
R}$ generators $\s^a$. Thus the mere switch in the spinor basis 
suffices to correctly implement the mirror map.

The fermion basis \eqn{ident} shows an obvious decomposition of 
the index $\a$ according to $\a = (I,r)$ with the index $r$ 
taking values $r=1,2$; a similar decomposition holds for $\bar 
\a$. This decomposition will be used below. For instance, the 
Sp(1)$\times$Sp($n$) one-forms,  can be written as $V^\a_{A\,i}\,
{\rm d}\phi^A=(V^I_A\,{\rm d}\phi^A)^r{}_i$.  Using \eqn{ident} 
we can now identify these one-forms as well as the Sp$(n)$ 
connections for a hypermultiplet theory that originates from a 
four-dimensional vector multiplet theory by comparing the fermion 
supersymmetry transformations on vector and hypermultiplet sides. 
We thus find (strictly speaking the indices $i$ now run over 
$+,-$),
\begin{equation}
V_{Ai}^\a \,{\rm d}\phi^A= \Big(V^I_A\,{\rm d}\phi^A\Big)^r{}_i= 
2\pmatrix{ {\rm d}X^I & N^{IK} \bar{\cal W}_K\cr
\noalign{\vskip 4mm}
-N^{IK}{\cal W}_K  & {\rm d}\bar X^I\cr}\,, \label{Vform}
\end{equation}
where ${\cal W}_I =  {\rm  d}B_I-  F_{IJ} {\rm d}A^J$ and 
$\a=(I,r)$, and  
\begin{equation}
\G_A{}^{\!\a}{}_{\!\b} \,{\rm d}\phi^A = 
\Big(\G_{A}\, {\rm d}\phi^A\Big)^{I\,r}{}_{\!J\,s} = \pmatrix{ -i 
N^{IK}\,{\rm d}F_{KJ} 
& -iN^{IK}\bar F_{KJL}N^{LM}{\cal W}_M    \cr
\noalign{\vskip 5mm} 
-iN^{IK}F_{KJL}N^{LM}\bar{\cal W}_M  &i N^{IK}\,{\rm d}\bar 
F_{KJ}\cr} \ ,
\end{equation}
with $\a=(I,r)$ and $\b=(J,s)$. Observe that the above quantities all take 
their values in the quaternions.

{}From the transformation rules and/or the action we can now 
determine all the relevant quantities in the hypermultiplet 
sector, such as the metric, the complex structures and the 
antisymmetric tensor $\O$. They are all consistent with the 
general results for hypermultiplets, derived in the previous section. 
Let us first give the expressions for the fermionic metric 
$G_{\bar \a\b}$ and the antisymmetric tensor $\O_{\bar\a\bar\b}$,  
\be
G_{\bar \a\b}= \ft14 N_{IJ} \,\d_{rs}\,, \qquad 
\O_{\bar\a\bar\b} =  \ft14 N_{IJ} \,\varepsilon_{rs}\,. 
\ee
The one-forms $\g^A$ take the form,
\begin{equation}
\gamma_{i\bar\a\,A} \,{\rm d}\phi^A = 
\Big(\gamma_{AI}\,{\rm d}\phi^A\Big)_{ri}=\ft{1}{2}\pmatrix{N_{IK}{\rm d}
X^K & \bar {\cal W}_I \cr 
\noalign{\vskip 4mm}
- {\cal W}_I & N_{IK}{\rm d}{\bar X}^K}\ ,    \label{gform}
\end{equation}
where $\bar\a=(r,I)$. Furthermore, we present the fermionic 
Lagrangian that follows from \eqn{3dvlagr} and \eqn{ident}, which 
exhibits most of the geometric quantities, such as the tensor $W$ 
defined in \eqn{defW}, 
\bea
  4\pi\,\lagr_{\rm {ferm}} &=&
    -\ft{1}{4} N_{IJ} \Bigl( \bar 
\zeta^{\overline{I1}}\pa\slash\zeta^{J1}
      + \bar \zeta^{\overline{I2}}\pa\slash\z^{J2}  + 
{\hbox{h.c.}} \Bigr)\nonumber\\
  & & +\ft{1}{4}i F_{IJK} \Bigl( \bar 
\zeta^{\overline{I1}}\pa\slash X^J \z^{K1} 
    -\bar \zeta^{\overline{I2}}\pa\slash X^J\z^{K2}  + \nonumber\\
  & & \qquad\quad\quad 2 N^{JL}\, \bar \zeta^{\overline{I2}}(\pa\slash B_L
    -F_{LM}\pa\slash A^M)\z^{K1}\Bigr) +{\hbox{h.c.}}\nonumber\\
  & & -\ft{1}{24}i\Bigl(F_{IJKL} + 3iN^{MN}F_{M(IJ}F_{KL)N} \Bigr)\,
    \bar \zeta^{\overline{I2}}\gamma_\mu\z^{J1}\, \bar 
   \zeta^{\overline{K2}}\gamma^\mu\z^{L1}  +{\hbox{h.c.}}\nonumber\\
  & & -\ft{1}{24} N^{MN}F_{MIJ}\bar{F}_{NKL}
    \Bigl( \bar \zeta^{\overline{K1}}\gamma_\mu\z^{I1} 
      -\bar \zeta^{\overline{I2}}\gamma_\mu\z^{K2} \Bigr)\nonumber\\
  & & \qquad\quad\quad\times \Bigl( \bar 
\zeta^{\overline{L1}}\gamma^\mu\z^{J1} 
      -\bar \zeta^{\overline{J2}}\gamma^\mu\z^{L2} \Bigr)\nonumber\\
  & & -\ft{1}{12} N^{MN} F_{MIJ}\bar{F}_{NKL}\,
    \bar \zeta^{\overline{I1}}\gamma_\mu\z^{J2}\, 
    \bar \zeta^{\overline{K1}}\gamma^\mu\z^{L2}\,. 
\eea
The tensor $W$ defined in \eqn{defW}, is thus expressed in terms 
of the tensor $C_{IJKL}$, defined in \eqn{defC}, and the 
curvature tensor of the special \Ka\ space given in \eqn{specialconcurv}. Both these 
tensors, and therefore the tensor $W$, are covariant with respect to the 
symplectic reparametrizations of the underlying special \Ka\ manifold.
The tensor $W$ fully encodes the 
curvature tensor of the special hyper-\Ka\ manifold. We refrain 
from giving explicit  
formulae, but wish to point out that these expressions allow for a 
coordinate-independent characterization of the special hyper-\Ka\ manifolds. 
We have also verified that the tensor $W$ becomes fully symmetric 
when written in purely (anti)holomorphic indices, employing the 
result for the tensors $G_{\bar \a\b}$ and $\O_{\bar\a\bar\b}$ 
given above.

It is clear from their index structure that the one-forms  
\eqn{Vform}  transform covariantly under the symplectic 
reparametrizations of the underlying vector multiplet by 
multiplication from the left with matrices
\be
S^{Ir}{}_{\! Js} =\pmatrix{ {\cal S}^I{}_J&0\cr 
\noalign{\vskip 2mm}  0& \bar{\cal S}^I{}_J}\,,
\ee
while the one-forms \eqn{gform} transform from the left with $[\bar 
S^{-1}]^J{}_I$. In general these transformations 
are not contained in the holonomy group Sp$(n)$.
              
The above thus constitutes the full construction of a 
hypermultiplet model in four spacetime dimensions associated with 
a specific theory based on vector multiplets. The detour through three 
dimensions only serves as a means to arrive at these results. 
Unlike the corresponding theory of vector multiplets, the 
hypermultiplet theory does not exhibit an SU$(2)_{\rm R}$ 
invariance, at least not in the generic case. Only a manifest 
U$(1)_{\rm R}$ invariance remains. All the isometries of the 
vector-multiplet target space that represent invariances of the 
full set of equations of motion, remain present as isometries of 
the hypermultiplet target space. The symplectic 
reparametrizations of the vector multiplets induce corresponding 
transformations on the hyper-\Ka\ side. In this way we deal with 
a large class of hyper-\Ka\ spaces. They can be expressed in 
terms of certain restrictions on the curvature tensor. 

We should stress that the general hypermultiplet action is 
encoded in the one-forms $V^\a_i$, but one has to provide one 
extra ingredient, such as  
the fermionic metric $G_{\bar \a\b}$, or the antisymmetric 
tensor $\O_{\bar\a\bar\b}$. The expressions given above for these 
quantities concern the special hyper-\Ka\ spaces and are given in 
special coordinates. As already alluded to earlier, it is 
straightforward to write them in a coordinate-independent way. In 
the case of local supersymmetry, the one-forms will become 
Sp(1) sections subject to an appropriate projective condition.

As a last application of the mirror map we turn to the 
possible central charges that can  
emerge in the supersymmetry algebra for a theory based on 
vector multiplets or hypermultiplets. As the symplectic 
reparametrizations can be performed in a supergravity background 
\cite{dWVP}, the algebra and therefore the expressions for the 
central charges should be invariant under these 
reparametrizations. Likewise, the charges should be consistent 
with the underlying \Ka\ or hyper-\Ka\ geometry. We will 
determine the central charges by evaluating the possible surface 
terms on the right-hand side of the anticommutator of two 
supercharges. To determine this anticommutator we use canonical 
quantization. This approach is the same as the one followed in
\cite{olive} for the elementary super-Yang-Mills system. Here we 
apply it for an arbitrary function $F(X)$ and arbitrary hyper-\Ka\ 
metrics. 

Let us first present the supercurrent for the vector multiplet and 
hypermultiplet theories,
\bea
J_{\mu \,i} &=& {1\over8\pi} \,\Big\{N_{IJ}\,\pa\slash\bar 
X^I\gamma_\mu\Omega_i^J + \ft12 i \varepsilon_{ij} \,
{\cal G}^-_{\rho\sigma\,I} \,\sigma^{\rho\sigma} 
\gamma_\mu\Omega^{jI} +\ft1{12}i \bar F_{IJK}\, \g_\m 
\O^{kI}\;\bar\O^{lJ}\O^{jK} \, \varepsilon_{ij}  \varepsilon_{kl} 
\Big\} \,,\nonumber\\    
J_{\mu \,i}&=& {1\over 4\pi}\, g_{AB}\,\g^A_{i\bar \a}\, 
\partial\slash\phi^B \g_\mu  \zeta^{\bar \a} \ . 
\eea
where ${\cal G}^-$ was defined in \eqn{defcG}. 
The other chirality components follow by complex conjugation. 
Observe that the first one is invariant under symplectic 
reparametrizations. Obviously the second expression for the 
hypermultiplet current is invariant under the hyper-\Ka\ holonomy 
group. The reader may be surprised that the vector-multiplet 
current contains terms cubic in the fermion fields, whereas the 
hypermultiplet current is linear in the fermion fields. 
Still 
one can verify, by performing the duality transformation in the presence of the 
gravitino field coupling to the supercurrent, that
the expressions for the two currents become 
compatible upon reduction to three dimensions. 

To determine the central charges one needs only the Dirac 
brackets for the fermions, as the bosonic brackets lead to 
terms at least quadratic in the fermion fields, which represent 
supersymmetric completions of bosonic terms that are  
already present in the algebra. In this way we find the following 
commutators for the vector multiplet,
\bea
\{Q_i,\bar Q^j\} &=& i\hbar\, {1-\gamma^5\over 2}\,\d_i{}^j \,
\Big\{\g_\m \,P^\m  + \g_a \,Z^a\Big\} \,, \nonumber\\
\{Q_i,\bar Q_j\} &=& -i\hbar\, (1-\gamma^5)\,
\varepsilon_{ij} \,\Big\{ \bar X^I\,q_{{\rm e}I}- \bar 
F_I\, q_{\rm m}^I \Big\}\,, \label{susyalg}
\eea
where the vector central charge, $Z^a$, is defined by ($a,b,c$ 
denote spatial indices), 
\be
Z^a =  {i\over 8\pi} \varepsilon^{abc}  \,\int {\rm d}^3x
\; N_{IJ}\, \partial_b X^I \,\pa_c \bar X^J \,, \label{Kflux}
\ee
which is an integral over the \Ka\ two-form; the second 
anticommutator yields the  
anti-holomorphic BPS mass expressed in terms of the values of 
$\bar X^I$ and $\bar F_I$ taken at spatial infinity (to obtain 
this result we used the field equations for the vector fields)  
and the electric and magnetic charges.\footnote{%
   The charges $q_{{\rm e}I}$ and $q_{\rm m}^I$ are related to 
   electric and magnetic charges and are defined in terms 
   of flux integrals over closed spatial surfaces that surround the 
   charged objects (quantized on a lattice with elementary area 
   equal to $2\hbar$),
\[
 2\pi \,q_{\rm m}^{I}= \oint_{\partial V} (F^+ + F^-)^I \,,
\qquad  
2\pi \,q_{{\rm e}I}= \oint_{\partial V} (G^+ + G^-)_I \,. 
\]
This definition shows that the charges $(q_{\rm m}^{I},q_{{\rm 
e}I})$ transform under symplectic reparametrizations precisely as 
the field strengths $(F^I,G_I)$. } %
Obviously the central charges are invariant under symplectic 
reparametrizations, as predicted above. 
For the case of a quadratic function $F$ our result for the 
second commutator coincides with that in \cite{olive}. The \Ka\ 
form contribution was presented in \cite{triestedw}.

For the hypermultiplets we find a similar result for the 
anticommutators,
\bea
\{Q_i,\bar Q^j\} &=& i\hbar \,{1-\gamma^5\over 2}\, \Big\{
\d_i{}^j \,\g_\m \, P^\m +  (\s^\L)_i{}^j \, \g_a \,Z^{\L\,a}  
\Big\} \,,  \nonumber\\ 
\{Q_i,\bar Q_j\} &=& 0 \,,
\eea
where we now have three vector central charges defined by
\be 
Z^{\L\,a} = - {1\over 16\pi}\varepsilon^{abc} \, \int {\rm d}^3 x 
\;  J^{\L}_{AB} \,\pa_b\phi^A\pa_c\, \phi^B \, . \label{HKflux}
\ee
The $J^\L$ are the three complex structures of the 
hyper-\Ka\ space defined in \eqn{cstructures}.

There is a clear systematics in the above results. Note that the 
central charges for the vector multiplet are singlets under 
SU$(2)_{\rm R}$, whereas those for the hypermultiplets transform 
as a triplet under this group. In addition 
to the BPS mass, we find certain integrals over the pull back of the 
\Ka\ form (for the vector multiplet) and the hyper-\Ka\ forms 
(for the hypermultiplet). Naively, all these 
integrals vanish, as we can write (locally in the target space) 
these two-forms as the exterior derivative of corresponding one-forms. This 
then allows us to write the integrands as total derivatives in 
the base space, which can be dropped subject to certain reasonable 
assumptions on the asymptotic values of the scalar fields.  
Hence the question whether these charges are actually realized 
depends on the kind of boundary conditions that one wishes to 
impose. For instance, in $3+1$ 
dimensions, if one imposes boundary conditions at spatial 
infinity such that the fields converge in all directions to the same 
value, with the derivatives vanishing sufficiently fast so as to 
ensure finite energy, then the central charges associated with 
the two-forms will vanish. 
In $2+1$ dimensions, the situation is different. In 
that case the central charges are expressed as integrals of the 
(hyper-)\Ka\ two-forms over the image of $\phi$. Topologically 
this image is $S^2$, so that the central charges are enumerated 
by the second homotopy group of the target-space manifold. Obviously 
the central charges set a BPS bound in the usual fashion. 

{}From the perspective of this paper it is of interest to see how 
the central charges of the vector multiplet sector and the 
hypermultiplet sector are related by mirror symmetry. When 
suppressing the dependence on the compactified coordinate $x^3$ 
the central charges $Z^3$ and $Z^{\L\,3}$ can be finite. It is 
then straightforward to write down the supersymmetry algebra 
corresponding to \eqn{susyalg} in three dimensions. One subtlety 
is that the momentum in the third direction is also a surface 
integral, which should be added to the central charge associated 
with the \Ka\ form. As it turns out, the resulting two-form 
corresponds then precisely with the \Ka\ form $\o^3$ defined in 
\eqn{complexstructures} for the hyper-\Ka\ space. 

In order to apply the mirror map, we write the charges in an  
alternative basis in correspondence with the new basis 
\eqn{newpara} for the supersymmetry parameters, 
\bea
Q^+=\ft12\sqrt{2} 
\g^3(Q_1-iQ_2)\,, \qquad
Q^- =\ft12\sqrt{2}(Q^1-iQ^2)\,, \nonumber \\
Q_+ =\ft12\sqrt{2} 
\g^3(Q^1+iQ^2)\,,\qquad
Q_- =\ft12\sqrt{2}(Q_1+iQ_2)\,.  \label{newQ}
\eea
With these definitions, the three-dimensional version of
\eqn{susyalg} reads
\bea
\{Q_\pm , \bar Q^\pm\} &=& i\hbar {1-\g^5\over 2} \,\Big\{\g_\m 
P^\m \mp i Z^\prime\Big\}\,,  \nonumber \\
\{Q_+ , \bar Q^- \} &=& -2i\hbar {1-\g^5\over 2} \,\Big\{ X^I\,
q_{{\rm e}I}- F_I\, q_{\rm m}^I \Big\}\,,  \nonumber \\
\{Q_- , \bar Q^+ \} &=& 2i\hbar {1-\g^5\over 2} \,\Big\{\bar 
X^I\,q_{{\rm e}I}- \bar  F_I\, q_{\rm m}^I \Big\}\,, 
\eea
where $Z^\prime$ is now defined in terms of the hyper-\Ka\ 
two-form $\o^3$.  
This result coincides with the algebra relevant to the 
hypermultiplets upon reduction to three spacetime dimensions, 
which reads,
\bea
\{Q_i,\bar Q^j\} &=& i\hbar \,{1-\gamma^5\over 2}\, \Big\{
\d_i{}^j \,\g_\m \, P^\m + i (\s^\L)_i{}^j \,  \,Z^{\L\,3}  
\Big\} \,,  \nonumber\\ 
\{Q_i,\bar Q_j\} &=& 0 \,.
\eea
This demonstrates that the supersymmetry algebra remains 
consistent with the mirror map in the presence of the central 
charge configurations. A gratifying feature of this result is 
that the holomorphic BPS mass of the vector multiplets is mapped 
to the holomorphic hyper-\Ka\ two-forms, $\o^\pm$, defined in 
\eqn{complexstructures}.

Although the above results do not capture the full 
dynamics of the four-dimensional gauge theories in a circle 
compactification, they are consistent with the results 
derived in the context of three-dimensional gauge 
dynamics \cite{3dmirror}. There the two sets of central charges 
are associated with  
explicit mass terms and Fayet-Iliopoulos terms, which are 
interchanged under the quantum mirror symmetry. The relation  
of the central charges with integrals of the 
hyper-\Ka\ two-forms also arose in that context. 
                                                          
\vspace{1cm}

{\bf Acknowledgement}\\[3mm]
We are grateful to J. de Boer and A. Van Proeyen for valuable 
discussions.  
J.~D.J. and S.~V. thank the Institute for Theoretical Physics of 
Utrecht University for hospitality.  

This work is supported by the European Commission TMR programme  
ERBFMRX-CT96-0045, in which J.~D.J. is  
associated to Leuven and B.~d.W. and B.~K. are associated
to Utrecht. 


\end{document}